\documentclass[12pt]{article}

\usepackage{epsfig}

\begin{document}
\renewcommand{\thefootnote}{\fnsymbol{footnote}}

\pagestyle{empty}
\pagenumbering{arabic}

\begin{flushright}
Revised Version\\
ZU-TH 30/96\\
TTP96--54\\
June 1997\\[15mm]
\end{flushright}

\begin{center}
{\Large Resonance Contributions to the Electromagnetic Low Energy 
Constants of Chiral Perturbation Theory}\\[15mm] 
Robert Baur\footnote{e-mail: rbaur@physik.unizh.ch}\\[2mm] 
Institut f\"ur Theoretische Physik, Universit\"at Z\"urich\\ 
CH-8057 Z\"urich, Switzerland\\[10mm] 
Res Urech\footnote{e-mail: ru@ttpux2.physik.uni-karlsruhe.de}\\[2mm] 
Institut f\"ur Theoretische Teilchenphysik, Universit\"at Karlsruhe\\ 
D-76128 Karlsruhe, Germany\\[18mm] 
{\Large {\bf Abstract}}\\
\end{center}
The effective chiral Lagrangian of the strong and electromagnetic
interactions of the pseudoscalar mesons at low energies depends on a set of
low energy constants. We determine the contributions to the electromagnetic
coupling constants at order $O(e^2 p^2)$, which arise from resonances
within a photon loop. We give some implications of our results, in
particular we discuss in detail the effects on the corrections to Dashen's
theorem.\\[5mm]
PACS number(s): 12.39.Fe, 13.40.Dk, 13.40.Ks, 14.40.Aq

\newpage
\pagestyle{plain}

\section{Introduction}\label{res:section1}

At low energies the strong and electromagnetic interactions of the lightest
pseudoscalar mesons can be described by an effective field theory, called
chiral perturbation theory (CHPT)
\cite{weinberg79,gasser84,leutwyler94,reviews}. CHPT provides a systematic
expansion in the external momenta $p$, in the light quark masses $m_q$ and,
in the electromagnetic sector, in the electromagnetic coupling $e$. It is a
nonrenormalizable field theory, but it is renormalized order by order in
the loop expansion. The effective chiral Lagrangian depends on a number of
low energy constants that absorb the ultraviolet divergences generated by
the loops. The renormalized couplings are not fixed by chiral symmetry.
They are in principle calculable from the underlying dynamics of QCD and
QED in terms of the renormalization group invariant scale $\Lambda$ and
the heavy quark masses $( m_{c},m_{b}, \ldots )$. In practice, however,
they are extracted from experimental data, which works to a large extent in
the strong sector but hardly in the electromagnetic sector \cite{ecker95}.

Complementary to experimental data on may use model assumptions as an
input. In the phenomenological picture, the low energy constants get
contributions from different sources, such as mesonic resonances, other
hadronic states, and even short distance effects. In \cite{ecker89} the
authors separated the low energy constants of the strongly interacting
sector at next-to-leading order in a part determined by resonance exchange,
and a non-resonant remainder. It was found that at the scale point
$\mu=M_\rho$ the resonance contribution nearly saturates the experimentally
known low energy parameters of the strong interaction sector.

In the electromagnetic sector, resonance exchange at tree level cannot be
applied because resonances do not contribute to the low energy
electromagnetic coupling constants in the absence of photons. Considering
the electromagnetic mass shift of the pion in the chiral limit due to
photon and resonance exchange at the one-loop level \cite{das67}, the
resonant contribution to the coupling constant $C$ at order $O(e^2)$ was
determined in \cite{ecker89} (see also \cite{bardeen89}). The occurring
divergence was absorbed in the corresponding electromagnetic counterterm
Lagrangian associated with the coupling constant $\hat C$. It was found
that the loop contribution is numerically very close to the experimental
value and the authors thus concluded that resonance saturation holds for
the leading order electromagnetic coupling constant $C$.

The purpose of this article is to extend the above method to the couplings
of the order $O(e^2 p^2)$. The effective electromagnetic Lagrangian has
been given in its general form at leading order in \cite{ecker89} and at
next-to-leading order in \cite{urech95,neufeld95,baur96b}. In the latter
case we have 14 independent operators associated with the coupling
constants $K_i\;(i=1\ldots 14)$. Considering the contributions to the
masses, to the scattering amplitudes, and to the matrix elements with
external currents (see \mbox{Section \ref{res:section4})} arising from
resonances within a photon loop, the resonant contributions to the coupling
constants are evaluated. The divergences generated by the loops must be
absorbed in a counterterm Lagrangian with the corresponding coupling
constants $\hat K_{i}$ \cite{baur96}. Afterwards we extract the finite
contributions from the resonances \mbox{to the $K_i$}.

Since our knowledge on the coupling constants of the electromagnetic sector
is still poor, i.e. in general they cannot be related to experimental data,
it is not possible to test whether resonance saturation holds in the
electromagnetic sector. We are aware of two possible checks. $(i)$ On the
one hand we can classify the operators at order $O(e^2 p^2)$ due to their
behaviour in the large $N_C$ limit \cite{hooft74} (see Appendix
\ref{appendixnc}) and compare them to the numerical result for the
couplings $K_i$ found in the resonance approach. In the four cases where a
suppression by $1/N_C$ occurs we find indeed that the contributions from
the resonances vanish, e.g. one of the results is $K_8^R=0$, where the
superscript $R$ indicates resonance contributions. $(ii)$ On the other
hand, exactly this coupling has been estimated in \cite{urech95} to be
$K^r_8(M_\rho) = - ( 4.0 \pm 1.7) \times 10^{-3}$, where the superscript
$r$ stands for the renormalized coefficient in the Lagrangian
${\cal{L}}^{Q}_{4}$ and the scale point is chosen to $\mu =
M_\rho$. Obviously, the two results do not coincide. So the only possible
hint for a resonance saturation for the $K_i$ is found in $(i)$.

We apply our results to different $S$-matrix elements given in the
literature (see Section \ref{res:section5}), e.g. we discuss the
consequences for the corrections to Dashen's theorem
\cite{dashen69,leutwyler95} at order $O(e^2 m_q)$. There we find a strong
cancellation among the contributions leading to the result that the
uncertainties are of the same size (or even larger) than the central value
itself.

This article is organized as follows: In Section \ref{res:section2} the
Lagrangians of chiral perturbation theory up to the orders $O(p^4)$ and
$O(e^2 p^2)$, respectively, are briefly presented.  
 In Section \ref{res:section3} the
linear couplings of the pseudoscalar mesons to the resonances and the
relations used among the parameters in the resonance sector are given.
Then the general
procedure in order to determine the resonance contributions to the low
energy coupling constants is described. In
Section \ref{res:section4} we present the evaluation of the resonant
contributions to the low energy coupling constants $K_i$, including an
extended discussion of the results. The applications of the numerical
results to $S$-matrix elements are given in Section \ref{res:section5} and
in the last Section a summary is presented. In Appendix \ref{appendixmass}
we discuss the heavy mass expansion, that we have used in the calculation
of the resonance-photon loops at order $O(e^2 p^2)$. In Appendix
\ref{appendixkr} we list in detail the contributions to the $K^R_i$
separated for each type of resonances and in Appendix \ref{appendixnc} the
large $N_C$ behaviour of the operators in the Lagrangian
${\cal{L}}^{Q}_{4}$ is deduced.

For comprehensive reviews on CHPT we refer the reader to \cite{reviews}, a
review on the determination of low energy couplings may be found in
\cite{ecker95}.

\section{Chiral Lagrangian and Low Energy Couplings}
\label{res:section2}

The chiral Lagrangian can be expanded in derivatives of the Goldstone
fields and in the masses of the three light quarks. The power counting is
established in the following way: The Goldstone fields are of order
$O(p^0)$, a derivative $\partial_\mu$, the vector and axial vector currents
$v_\mu, a_\mu$ count as quantities of $O(p)$ and the scalar (incorporating
the masses) and pseudoscalar currents $s,p$ are of order $O(p^2)$. The
effective Lagrangian starts at $O(p^2)$, and is  denoted by ${\cal L}^{Q}_{2}$. It is the non-linear $\sigma$-model Lagrangian coupled to external fields,
respects chiral symmetry $SU(3)_{R}\times SU(3)_{L}$, and is invariant
under $P$ and $C$ transformations \cite{gasser84}. The Lagrangian ${\cal
L}^{Q}_{2}$ including electromagnetic interactions can be divided in three
parts, namely \cite{ecker89}
\begin{equation}
{\cal{L}}^{Q}_{2} = {\cal{L}}^{\gamma}_{kin} + {\cal{L}}_{2} + 
{\cal{L}}^{C}\;,
\label{lq2}
\end{equation}
where 
\begin{eqnarray}\label{l2}
{\cal{L}}^{\gamma}_{kin} & = & - \frac{1}{4} F_{\mu \nu} F^{\mu \nu} 
                              - \frac{\lambda}{2}
                              (\partial_{\mu} A^{\mu})^{2} \nonumber \\ 
{\cal{L}}_{2} & = & \frac{F^{2}_{0}}{4}\langle 
                   d_{\mu}U d^{\mu}U^{\dagger} \rangle 
                   + \frac{F^{2}_{0}}{4}\langle \chi^{\dagger} U 
                   + U^{\dagger} \chi \rangle \nonumber \\
{\cal{L}}^{C} & = & C\langle QUQ U^{\dagger} \rangle \; .
\end{eqnarray}
Here, $F_{\mu \nu}$ is the field strength tensor of the photon field
$A_{\mu}$. The parameter $\lambda$ is the gauge fixing parameter, which is
set to $\lambda = 1 $ henceforth. The pseudoscalar meson fields $\phi_a$
are contained in the usual way in the matrix $U=\mbox{ exp}\,(i/F_0 \cdot
\sum_a \lambda_a \phi_a)$, and the field $\chi$ incorporates the scalar and
pseudoscalar currents $s$ and $p$ respectively, $\chi = 2 B_{0} (
s+ip)$. The quark mass matrix is contained in $s$ (i.e. $s = {\cal M} +
\cdots$), and we work in the isospin limit $m_u = m_d = \hat{m}$,
i.e. ${\cal M} = \mbox{ diag}\,(\hat{m},\hat{m},m_s)$.  The covariant
derivative $ d_{\mu}U$ defines the coupling of the pseudoscalar mesons to
the photon field $A_{\mu}$, the external vector and axial vector currents
$v_{\mu}$ and $a_{\mu}$ respectively,
\begin{equation}
         d_{\mu}U = \partial_{\mu} U - i(v_{\mu}+Q A_{\mu} +a_{\mu}) U
                                     + iU(v_{\mu}+Q A_{\mu} -a_{\mu})\;,
\end{equation}
where the charge matrix $Q$ is given by $Q = e\mbox{
diag}\,(2/3,-1/3,-1/3)$. $F_{0}$ corresponds to the pion decay constant
$F_\pi$ in the chiral limit and $B_0$ is related to the quark condensate
for $m_q \to 0$.

In the chiral limit the masses of the Goldstone bosons are of purely
electromagnetic nature, determined by the operator proportional to $C$, 
\begin{equation}
        C\langle QUQ U^{\dagger} \rangle =
-\frac{2e^2 C}{F^{2}_{0}}(\pi^{+}\pi^{-}+K^{+}K^{-}) + O(\phi^4).
\label{cterm}
\end{equation}
The masses of the charged particles receive an overall shift, whereas the
masses of the neutral fields remain zero, in agreement with Dashen's
theorem \cite{dashen69}. 

In order to formally maintain a consistent 
chiral counting, it is convenient to \mbox{set \cite{urech95}}
\begin{eqnarray}
         e \sim O(p), \hspace{5mm} A_{\mu}  \sim O(1)\;,  
\end{eqnarray}
such that ${\cal L}^{Q}_{2}$ is of the order $O(p^2)$.
 
At next-to-leading order the most general chiral invariant, P and C
symmetric Lagrangian ${\cal L}^Q_4$ at order $O(p^4)$ has been given in
\cite{gasser84} for the strong interaction (associated with the couplings
$L_i$ and $H_i$) and in \cite{urech95,neufeld95,baur96b} for the
electromagnetic interaction (couplings $K_i$),
\begin{eqnarray}\label{lag4}
{\cal{L}}^{Q}_{4}& = & L_{1} \langle d^{\mu}U^{\dagger } 
                                     d_{\mu}U \rangle^{2} 
   + L_{2} \langle d^{\mu}U^{\dagger } d^{\nu} U \rangle 
           \langle d_{\mu}U^{\dagger } d_{\nu}U \rangle \nonumber \\ 
&& + L_{3} \langle d^{\mu}U^{\dagger } d_{\mu}U d^{\nu}U^{\dagger } 
                   d_{\nu}U \rangle 
+ L_{4}\langle d^{\mu}U^{\dagger } d_{\mu}U \rangle 
       \langle \chi^{\dagger} U + U^{\dagger} \chi \rangle \nonumber \\ 
&& + L_{5} \langle d^{\mu}U^{\dagger } d_{\mu}U \left( \chi^{\dagger} U 
           + U^{\dagger} \chi \right) \rangle 
   + L_{6} \langle \chi^{\dagger} U + U^{\dagger} \chi \rangle ^{2} 
   + L_{7} \langle \chi^{\dagger} U - U^{\dagger} \chi \rangle ^{2} 
\nonumber \\ 
&& + L_{8} \langle \chi^{\dagger} U \chi^{\dagger} U + \chi U^{\dagger}
           \chi U^{\dagger} \rangle 
   - i L_{9} \langle d^{\mu}U d^{\nu}U^{\dagger } F_{R\mu \nu} 
             + d^{\mu}U^{\dagger} d^{\nu} U F_{L\mu \nu} \rangle 
\nonumber \\ 
&& + L_{10} \langle U^{\dagger} F^{ \mu \nu}_{R} U F_{L\mu \nu} \rangle 
   + H_{1} \langle F^{ \mu \nu}_{R} F_{R\mu \nu} 
           + F^{ \mu\nu}_{L} F_{L\mu \nu} \rangle 
   + H_{2} \langle \chi^{\dagger} \chi \rangle \nonumber \\ 
&& + K_{1} F^{2}_{0} \langle d^{\mu}U^{\dagger } d_{\mu}U \rangle 
                    \langle Q^{2} \rangle 
   + K_{2} F^{2}_{0} \langle d^{\mu}U^{\dagger } d_{\mu}U \rangle 
                     \langle Q U Q U^{\dagger} \rangle \nonumber \\ 
&& + K_{3} F^{2}_{0} \left( \langle d^{\mu}U^{\dagger } Q U \rangle 
                     \langle d_{\mu}U^{\dagger } Q U \rangle 
                     + \langle d^{\mu}U Q U^{\dagger} \rangle 
                       \langle d_{\mu}U Q U^{\dagger} \rangle \right) 
\nonumber \\ 
&& + K_{4} F^{2}_{0} \langle d^{\mu}U^{\dagger } Q U \rangle 
                     \langle d_{\mu}U Q U^{\dagger} \rangle 
   + K_{5} F^{2}_{0} \langle \left\{ d^{\mu}U^{\dagger }, d_{\mu}U \right\} 
                  Q^2 \rangle \nonumber \\
&& + K_{6} F^{2}_{0} \langle d^{\mu}U^{\dagger} d_{\mu}U Q U^{\dagger} Q U 
                     + d^{\mu}U d_{\mu}U^{\dagger } Q U Q U^{\dagger}
                     \rangle  \nonumber \\ 
&& + K_{7} F^{2}_{0} \langle \chi^{\dagger} U + U^{\dagger} \chi \rangle 
                     \langle Q^2 \rangle 
   + K_{8} F^{2}_{0} \langle \chi^{\dagger} U + U^{\dagger} \chi \rangle
                     \langle Q U Q U^{\dagger} \rangle \nonumber \\ 
&& + K_{9} F^{2}_{0} \langle \left( \chi^{\dagger} U + U^{\dagger} \chi
                     \right) Q^2 +\left( \chi U^{\dagger}
                     + U \chi^{\dagger} \right) Q^2 \rangle \nonumber \\ 
&& + K_{10} F^{2}_{0} \langle \left( \chi^{\dagger} U + U^{\dagger} \chi 
                     \right) Q U^{\dagger} Q U + \left( \chi U^{\dagger} 
                     + U \chi^{\dagger} \right) Q U Q U^{\dagger} \rangle
\nonumber \\ 
&& + K_{11} F^{2}_{0} \langle \left( \chi^{\dagger} U - U^{\dagger} \chi 
                     \right) Q U^{\dagger} Q U + \left( \chi U^{\dagger} 
                     - U \chi^{\dagger} \right) Q U Q U^{\dagger} \rangle
\nonumber \\ 
&& + K_{12} F^{2}_{0} \langle d_{\mu} U^{\dagger } \left[ c_R^\mu Q,Q \right]
                     U + d_{\mu}U \left[ c_L^\mu Q, Q \right] 
                     U^{\dagger} \rangle \nonumber \\[2mm] 
&& + K_{13}F^{2}_{0} \langle c_R^{\mu} Q U c_{L\,\mu} Q U^{\dagger} \rangle 
   + K_{14}F^{2}_{0} \langle c_R^{\mu} Q  c_{R\,\mu} Q + c_L^{\mu} Q  
                  c_{L\,\mu} Q \rangle \nonumber \\[2mm] 
&& + K_{15} F^{4}_{0} \langle Q U Q U^{\dagger} \rangle ^2 
   + K_{16} F^{4}_{0} \langle Q U Q U^{\dagger} \rangle \langle Q^2 \rangle 
   + K_{17} F^{4}_{0} \langle Q^2 \rangle^{2} \;,
\end{eqnarray}  
where 
\begin{eqnarray}
F_{R,L}^{\mu} & = & v^{\mu}+Q A^{\mu} \pm a^{\mu} \nonumber \\ 
F_{R,L}^{\mu \nu} & = &
   \partial^{\mu} F_{R,L}^{\nu} - \partial^{\nu} F_{R,L}^{\mu} 
   - i\left[ F_{R,L}^{\mu}, F_{R,L}^{\nu} \right] \nonumber \\
c_{R,L}^\mu & = & i\left[Q,F_{R,L}^{\mu}\right]\;.
\end{eqnarray}
$L_{1} \ldots L_{10},H_1,H_2$ and $K_{1} \ldots K_{17}$ are real low energy
constants, which are independent of the Goldstone bosons masses and which
parameterize all the underlying physics (including e.g. the resonances). The
coupling constants $H_{1},H_{2},K_{14}$ and $K_{17}$ have no physical
significance and are needed for renormalization only.
The  constants $K_{1},K_{7}$ and $K_{16}$ are electromagnetic corrections to
$F_{0}$, $B_{0}$, and $C$ respectively. Considering $n$-point functions
that involve vector and axial vector currents only, the coupling constant
$K_9$ can be absorbed in the quark mass matrix,
\begin{equation}
{\cal M} \longrightarrow {\cal M}^Q = {\cal M} 
                         \left( {\bf 1} + 8 K_9 Q^2 \right) \quad .
\end{equation}
Note that this substitution can be performed in particular in the
corrections to the meson masses. The operators associated with the
couplings $K_{15},K_{16}$ and $K_{17}$ are of the order $O(e^4)$ and are
not considered in the following.

\section{Resonances in the Chiral Lagrangian} \label{res:section3}

In order to describe the chiral couplings of the resonances to the
pseudoscalar Goldstone bosons, we restrict ourselves to the leading order
Lagrangian that is linear in the resonance fields. This Lagrangian contains
the kinetic term with a covariant derivative and the linear couplings of
resonances to the Goldstone bosons and to the external currents in such a
way, that all the terms involved are of order $O(p^2)$ \cite{ecker89}. We
do not include couplings bilinear in the resonances, (for a systematic
treatment in the heavy meson formalism see \cite{jenkins95}), nor consider
terms linear in the resonances at higher orders in $p^2$. Explicitly, the
interaction Lagrangian has the form \cite{ecker89}
\begin{eqnarray}\label{lr}
{\cal L}^V_2 & = & \frac{F_V}{2\sqrt{2}}\langle V_{\mu\nu} 
                   f_+^{\mu\nu}\rangle + \frac{iG_V}{2\sqrt{2}}\langle
                   V_{\mu\nu}[u^\mu,u^\nu]\rangle \nonumber \\
{\cal L}^A_2 & = & \frac{F_A}{2\sqrt{2}}\langle A_{\mu\nu} 
                   f_-^{\mu\nu}\rangle \nonumber \\
{\cal L}^S_2 & = & c_{d} \langle S u_{\mu} u^{\mu}\rangle 
                   + c_{m} \langle S \chi_{+} \rangle
                   +\tilde{c}_d S_{1} \langle  u_{\mu} u^{\mu}\rangle
                   +\tilde{c}_m S_{1}\langle  \chi_{+}\rangle \; ,
\end{eqnarray}
where
\begin{eqnarray}
f_\pm^{\mu\nu} & = & u F^{\mu\nu}_L u^\dagger \pm u^\dagger
                     F^{\mu\nu}_R u \nonumber \\
\chi_+ & = & u^{\dagger} \chi u^{\dagger} + u \chi^{\dagger} u
\nonumber \\
u_\mu & = & i u^\dagger d_\mu U u^\dagger =u^{\dagger}_{\mu} \nonumber \\
U & = & u^2 \; .
\end{eqnarray}
For the description of the vector $(V)$ and axial vector $(A)$ resonances
the antisymmetric tensor notation is used. At leading order, for $V$ and
$A$ only the octets couple to the Goldstone bosons whereas for the scalar
resonances both the octet $(S)$ and the singlet $(S_1)$ contribute. There
exist also similar couplings for the excited pseudoscalar octet $(P)$ and
the singlet $(P_1)$, but since they do not generate electromagnetic
contributions at the order $O(e^2 p^2)$ we have omitted the Lagrangian
${\cal L}^P_2$ in Eq.(\ref{lr}). The coupling constants occurring in
(\ref{lr}) will be discussed below.

In the kinetic Lagrangian a covariant derivative acts on the octet fields,
\begin{eqnarray}\label{rkin}
{\cal L}^{V,A}_{kin} & = & -\frac{1}{2}\langle \nabla^\mu R_{\mu\nu}
                       \nabla_\sigma R^{\sigma\nu} - \frac{1}{2}M^2_{R}
                       R_{\mu\nu}R^{\mu\nu}\rangle \nonumber \\
{\cal L}^{S}_{kin} & = & \frac{1}{2}\langle \nabla^\mu S 
                         \nabla_\mu S -M^2_{S}R^2\rangle 
                         + \frac{1}{2} \left(\partial^{\mu} 
                         S_{1}\partial_{\mu} S_{1} 
                         -  M^2_{S_{1}} S^2_{1}\right) \nonumber \\[2mm]
\nabla^\mu R_{(\mu\nu)} & = & \partial^\mu R_{(\mu\nu)} 
                              + [\Gamma^\mu,R_{(\mu\nu)}] \nonumber \\
\Gamma^\mu & = & \frac{1}{2}\left[ u^\dagger \left( \partial^\mu
                 - i F_R^\mu \right) u + u \left( \partial^\mu 
                 - i F_L^\mu \right) u^\dagger \right]\;,
\end{eqnarray}
where $M_R$ and $M_{R_1}$ are the corresponding masses in the chiral
limit. In the following we will calculate matrix elements containing
resonances within a photon loop. These loops contain divergences that can
be absorbed in a counterterm Lagrangian ${\cal L}_{c.t.}$. In its general
form, this Lagrangian has one term of order $O(e^2)$ and 14 terms of $O(e^2
p^2)$
\begin{equation} 
{\cal L}_{c.t.} = \hat C\langle Q U Q U^\dagger\rangle + 
                        \sum^{14}_{i=1} \hat K_i O_i.
\end{equation}
The operators $O_i$ are identical to those in Eq.(\ref{lag4}). The
necessary set of Lagrangians for the calculation of processes with
resonance-photon loops is given by
\begin{equation}
{\cal L}^R =  {\cal{L}}^{\gamma}_{kin} + {\cal L}^R_{kin} + {\cal L}^R_2  
              + {\cal L}_2 +{\cal L}_{c.t.}\hspace{2cm}R=V,A,S \,.
\end{equation}
The chiral power counting normally used in the loop expansion of the
Goldstone bosons cannot been maintained in the resonance sector. The masses
of the resonances do not vanish in the chiral limit, i.e. they have to be
considered as a quantity of the order $O(1)$. We therefore expand the loops
in the ratio $M_{\pi,\,K}^2/M_R^2$ and truncate at the order in
question. The procedure we use for this purpose is called the heavy mass
expansion, see Appendix \ref{appendixmass} for details and examples, where
we also discuss  the higher order corrections.

The coupling constants in Eq.(\ref{lr}) are not restricted by chiral
symmetry, however, they can be related to each other invoking the
asymptotic behaviour of QCD and by using large $N_C$ arguments
\cite{ecker89},
\begin{eqnarray} \label{sumrules}
\begin{array}{ll}
\mbox{   Relation}           & \mbox{      Source}             \\
\hline \\
F_A^2 = F_V^2 - F_0^2        & \mbox{First Weinberg sum rule
\cite{weinberg67}} \\[2mm]
M_A^2 F_A^2 = M_V^2 F_V^2\hspace{5mm} & \mbox{Second Weinberg sum rule
\cite{weinberg67}} \\[2mm]
F_V G_V = F_0^2              & \mbox{Electromagnetic pion form factor} 
\\[2mm]
F_V = 2 G_V                  & \mbox{Axial pion form factor} \\[4mm]
\left.\hspace{-2mm}
\begin{array}{l}
M_{S_1} = M_S \\[4mm] 
\displaystyle |\tilde{c}_d| = \frac{|c_d|}{\sqrt{3}} \\[4mm]              
\displaystyle |\tilde{c}_m| = \frac{|c_m|}{\sqrt{3}}
\end{array}
\right\} & \mbox{Large }N_C \mbox{ arguments}
\end{array}
\end{eqnarray}
Note that these relations are valid in the chiral limit only, with the
exception of the first Weinberg sum rule that also converges in the
presence of quark masses \cite{floratos79}. On the basis of
(\ref{sumrules}) all the parameters in the vector and axial vector sector
may be expressed as functions of the pion decay constant $F_0 \simeq F_\pi
= 92.4 \mbox{ MeV}$ and the rho mass $M_V \simeq M_\rho = 770\mbox{ MeV}$,
\begin{eqnarray}\label{values}
\begin{array}{lll}
|F_V| = \sqrt{2} F_0\;,\hspace{1cm} & 
\displaystyle |G_V| = \frac{F_0}{\sqrt{2}} \;, 
\hspace{1cm} & F_V G_V 0\;, \\[4mm]
|F_A| = F_0\;,                      & M_A = \sqrt{2} M_V\;. &
\end{array}
\end{eqnarray}
For a confirmation of the positivity of the product $F_V G_V$ from
experimental data we refer to \cite{neufeld95b}.
 
In the scalar sector more information is needed to fix $c_d$ and
$c_m$. Following the reasoning in \cite{ecker89} one may turn around the
arguments for the scalar contributions in the strong sector (see also the
footnote above): the two low energy constants $L_5$ and $L_8$ are by
assumption determined by the resonances, i.e. the finite (experimental)
values $L_5^r(\mu),L_8^r(\mu)$ fix the couplings $c_d,c_m$. Therefore at
$\mu = M_\rho$,
\begin{equation}
|c_d| = M_S \frac{L_5^r(M_\rho)}{\sqrt{2 L_8^r(M_\rho)}}, \hspace{2cm}
|c_m| = M_S \sqrt{2 L_8^r(M_\rho)}\;,
\end{equation}
and with the numerical values (where we omit any errors) $M_S = M_{a_0} =
982 \mbox{ MeV}$, $L_5^r(M_\rho) = 1.4 \times 10^{-3}$ and $L_8^r(M_\rho) =
0.9 \times 10^{-3}$ we arrive at
\begin{equation}\label{scalarvalues}    
|c_d| = 32 \mbox{ MeV}, \hspace{2cm} |c_m| = 42 \mbox{ MeV}, \hspace{2cm}
c_d c_m 0\;.
\end{equation}
For a different set of values for $c_d,c_m$ (due to a different
$L_8^r(\mu)$) see \cite{bramon94}. Note that we will need in our calculation
$c_d,\tilde{c}_d$ only, the couplings of the scalar mesons to the quark
masses yields higher order electromagnetic corrections.

For the masses of the pseudoscalar bosons we will use $M_\pi = 135 \mbox{
MeV}$, \mbox{$M_K = 495 \mbox{ MeV}$} and $M_\eta = 547 \mbox{ MeV}$.

\section{General Procedure}

The low energy coupling constants, introduced in the Chiral Lagrangian
${\cal L}^Q_4$, see (\ref{lag4}), are in general divergent. They absorb the
divergences at the one-loop level of the generating functional associated
with ${\cal{L}}^{Q}_{2}$
\cite{gasser84,urech95,neufeld95,baur96b}. Consequently, they have to be
renormalized and the renormalized coupling constants, denoted by
$L^{r}_{i}(\mu)$ and $K^{r}_{i}(\mu)$, will depend on the renormalization
scale $\mu$.

All low energy parameters of the chiral Lagrangian are determined by the
underlying dynamics of QCD and QED. Presently it is not possible to
calculate them from first principle. Referring, however, to data from low
energy experiments and large $N_{C}$ arguments the values of the
$L^{r}_{i}(\mu)$ have been determined in \cite{gasser84}. This is in
contrast to the electromagnetic sector, where only an estimate for
$K^{r}_{8}(\mu)$ has been given \cite{urech95}, and the determination of the
leading order coupling constant $C$ is based on resonance saturation
\cite{ecker89}.

In order to study the resonant contributions to the $L^{r}_{i}(\mu)$, Ecker
{\it et al.} \cite{ecker89} separated the low energy coupling constants at
a fixed scale point $\mu$ into a sum of resonant $(L^{R}_{i})$ and
non-resonant $(\hat L_{i}(\mu)\,)$ contributions, 
\begin{equation}\label{lires}
L^{r}_{i}(\mu)=\sum_{R= V,A,S,P} L^{R}_{i} + \hat L_{i}(\mu)\;,
\end{equation}
where the sum extends over vector (V), axial vector (A), scalar (S), and
pseudoscalar (P) resonances\footnote{Recently, the contributions arising
from tensor mesons have been determined in CHPT with two light flavours
\cite{toublan96}.}. Since the resonance exchange occurs at tree level, the
corresponding terms $L^{R}_{i}$ are scale independent. The choice of the
renormalization scale $\mu$ is not fixed a priori, however, in order to
best visualize the contributions from the resonances, $\mu$ should be in
the relevant mass region. Indeed, at $\mu= M_\rho$ it was found that the
finite parts of the strong coupling constants $L_i$ are almost determined
by resonance exchange, i.e. $\hat L_{i}(M_\rho) \simeq 0$
\cite{ecker89}\footnote{Exceptions to this rule are $L_5$ and $L_8$, where
scalar resonances contribute only. Due to the lack of experimental
information in this sector, the resonance saturation hypothesis was
assumed, finding good agreement with the decay width $\Gamma(a_0 \to \pi
\eta)$ \cite{ecker89}.}. To be more specific, whenever vector resonances
contribute at all, they dominate numerically (for a similar finding see
also \cite {donoghue89}). 

The same ansatz has been used in the electromagnetic sector for the leading
order coupling $C$. In contrast to the strong sector resonances contribute
to $C$ only within a photon loop, thus the resonant part will be divergent
in general. These divergences are absorbed by renormalizing the
corresponding non-resonant coupling $\hat C$ (which takes the role of a
counterterm) and we arrive at the splitting
\begin{equation}\label{ansatz1}
C  =\sum_{R} C^{R}(\mu_0)+  \hat C^{r}(\mu_0) \, ,
\end{equation}
where $C^{R}(\mu_0)$ and $\hat C^{r}(\mu_0)$ are finite and the scale
dependence cancels in the sum.

To be more explicit we take the electromagnetic mass shift of the charged
pion at the order $O(e^2)$ as an example. In ordinary CHPT without
resonances the charged pion mass reads in the chiral limit (see
Eq.(\ref{cterm}) above)
\begin{equation}
(M_{\pi^\pm}^2)_{e.m.} = \frac{2e^2}{F_0^2}C + O(e^2m_q)
\end{equation}
entirely determined by the coupling constant $C$. In the resonance
approach the pion mass also gets contributions from resonance-photon loops
at order $O(e^2)$, see graphs c) and d) in Figure \ref{appd:fig1} in
Appendix \ref{appendixmass},
\begin{eqnarray}
(M_{\pi^\pm}^2)_{e.m.} =   & - &\frac{3e^2}{F^2_0 16\pi^2}
                       \left[ F^2_V M^2_V \left(\ln\frac{M^2_V}{\mu^2}
                      + \frac{2}{3}\right)
                      -F^2_A M^2_A\left(\ln\frac{M^2_A}{\mu^2}
                      + \frac{2}{3}\right)\right] \nonumber \\[2mm]
                     &+&\frac{2e^2\hat{C}}{F^2_0} - \frac{6e^2}{F^2_0}
                       (F^2_V M^2_V - F^2_A M^2_A)\lambda 
                      + O(e^2m_q) \,  . 
\end{eqnarray}
The term $\lambda$ is given in Eq.(\ref{ms}) in Appendix
\ref{appendixmass}. It contains the loop divergences and produces a pole in
$d=4$ dimensions. It can be absorbed by renormalizing the contributions
from the non-resonant part,
\begin{equation}
\hat C = \hat C^r (\mu_0)+3(F^{2}_{V}M^{2}_V-F^{2}_{A}M^{2}_{A})\lambda \, .
\label{div1}
\end{equation}
Invoking Weinberg sum rules \cite{weinberg67}, see Eq.(\ref{sumrules}), one
circumvents the renormalization of $\hat C$, because the loop divergences
cancel. The connection to ordinary CHPT without resonances is then given by
the relation (\ref{ansatz1}) with the explicit expression for the
resonance contributions,
\begin{equation}\label{cr}
 C^R=  -\frac{3}{2}\frac{F_V^2 M_V^2}{16\pi^2}
                     \left(\ln \frac{M_V^2}{M_A^2}  \right) \;,
\end{equation}
where again we have used Weinberg sum rules. Putting numbers in (\ref{cr})
and comparing with the observed mass difference $M_{\pi^\pm}^2 -
M_{\pi^0}^2$ one finds that the resonances nearly saturate the coupling $C$
\cite{ecker89,das67}.

For the couplings $K_i$ we proceed in a completely analogous way. We split
the finite part $K^r_i(\mu)$ at a fixed scale point $\mu$ in a resonant
part and a remainder. The contribution from the resonances $K^R_i(\mu_0)$
depends on the renormalization scale $\mu_0$ at which one subtracts the
occurring divergences in the photon-resonance loops,
\begin{equation}\label{ansatz2}
K^{r}_{i}(\mu)=\sum_{R} K^{R}_{i}(\mu_0) +  \hat  K^{r}_{i}(\mu,\mu_0) \;,
\end{equation}
where each term is finite and the $\mu_0$ dependence cancels in the sum on
the r.h.s. Again, by the use of sum rules we find several combinations
of $K^R_i$ that are finite and thus $\mu_0$ independent. But in general,
measurable quantities (like the charged pion mass $M_{\pi^\pm}^2$) involve
combinations of $K^R_i$ that do not show this nice feature and therefore
the resonant contributions to these observables remain $\mu_0$ dependent
\cite{baur96}.

Similarly to the strong sector, we expect the coupling constants
$K^r_i(\mu)$ to be most sensitive to the contributions from the resonances
in the proximity of the lightest resonance $\rho$, i.e. we adopt the scale
point $\mu = M_{\rho}$ henceforth. Concerning the renormalization scale
$\mu_0$ in the photon-resonance loops we choose as central value $\mu_0 =
M_{\rho}$ and vary it in the region $M_\eta \sim 1\;{\rm GeV}$
in order to study the scale dependence of the couplings $K^R_i(\mu_0)$. 

In the strong sector and for the leading order electromagnetic coupling $C$
it was found, that there is not much room left for other contributions
beside the meson resonances. At $O(e^2p^2)$ we have poor informations only
(see Section \ref{res:section4}) which would establish a similar
finding. For the time being, we will treat the finite parts of the
renormalized remainder $\hat{K}^r_i(\mu, \mu_0)$ as unknown parameters.

\section{Resonance Contributions to the $K^r_i(\mu)$} 
\label{res:section4}

In order to determine the resonant contributions to the electromagnetic
couplings $K_i$, we have calculated $S$-matrix elements with the Lagrangian
${\cal L}^R$ at the one-loop level and have evaluated the corresponding tree
contributions from ${\cal L}^Q_4$. We then   split the coupling constants
$K_i$ according to Eq.(\ref{ansatz2}) in a resonant piece ($K^R_i$) and a
remainder ($\hat{K}_i$). The divergences occurring in the resonant part of
the loops are canceled by renormalizing the $\hat{K}_i$. The finite part
of the resonant loops are finally identified with the corresponding $K^R_i$
at the scale point $\mu_0 =M_\rho$.

\begin{figure}[p]
\begin{center}
\leavevmode
\epsfig{file=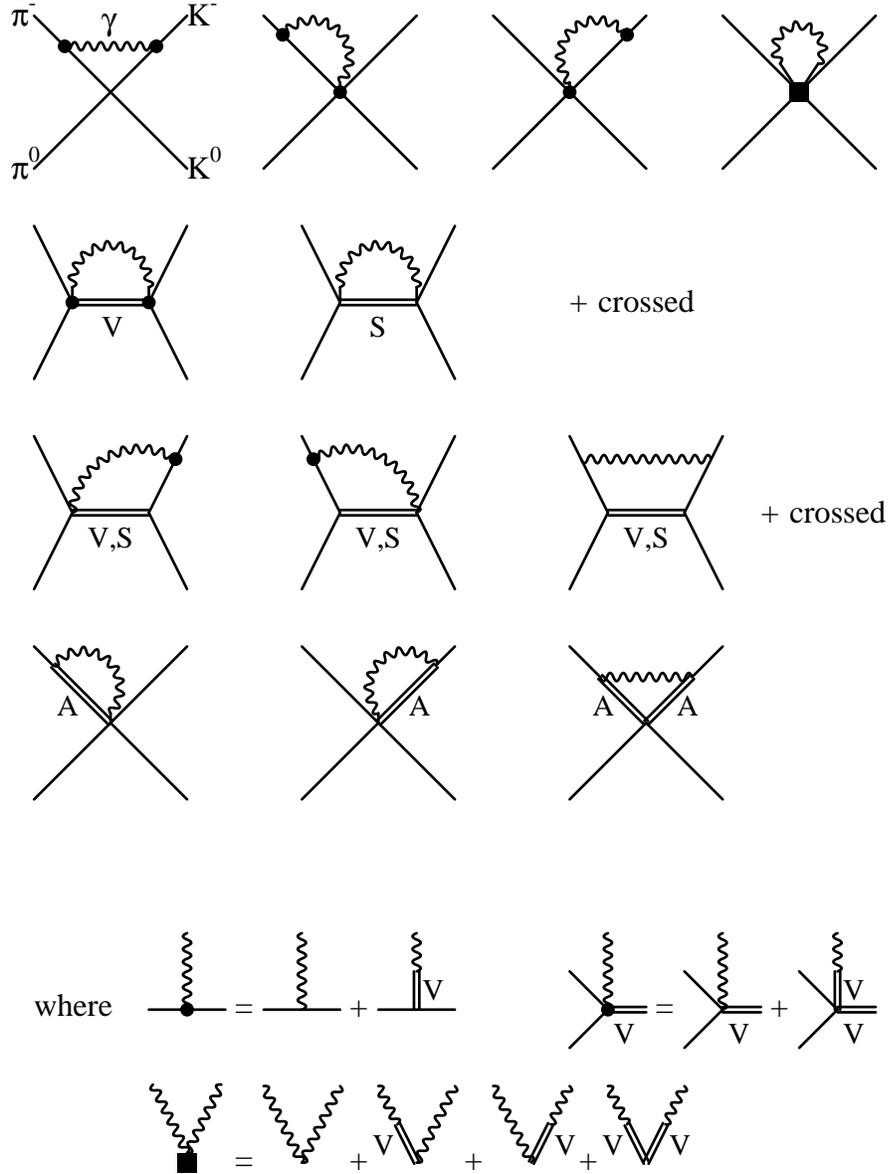,width=12cm}
\caption[]{\label{res:figmatch}
One-loop contributions to the scattering amplitude of $\pi^-
\pi^0 \to K^- K^0$. The double lines represent the resonances, the letter
nearby indicates their type: vector $(V)$, axial vector $(A)$ or scalar
$(S)$. Note that the vertices with two resonances are induced by the
covariant derivative in (\ref{rkin}).} 
\end{center}
\end{figure}

We have calculated the following set of $S$-matrix elements at the order
$O(e^2 p^2)$:
\begin{itemize}
\item Masses 
\begin{equation}
M_{\pi^\pm}, \hspace{3mm} M_{\pi^0}, \hspace{3mm} M_{K^\pm}, \hspace{3mm} 
M_{K^0}, \hspace{3mm}M_\eta\;.
\end{equation}
\item Scattering Amplitudes (in the chiral limit)
\begin{equation}
\pi^+ \pi^- \to \pi^+ \pi^-,\hspace{3mm} 
\pi^+ \pi^- \to K^0 \bar{K}^0,\hspace{3mm}
\pi^- \pi^0 \to K^- K^0 \;.         
\end{equation}
\item Matrix Elements (in the chiral limit) 
\begin{eqnarray}
&&\langle \pi^- | A_\mu^{1-i2}(0) | 0 \rangle,\hspace{3mm}
\langle \pi^+ | P^{1+i2}(0) | 0 \rangle,\hspace{3mm} 
\langle \pi^- | V_\mu^{4-i5}(0) | K^0 \rangle, \nonumber \\
&&\langle 0 | T\; V_\mu^{1-i2}(x) V_\nu^{1+i2}(y) | 0 \rangle,\hspace{3mm}
\langle 0 | T\; A_\mu^{1-i2}(x) A_\nu^{1+i2}(y) | 0 \rangle\;,
\end{eqnarray}
where
\begin{eqnarray}\label{current}
X^{k + il}(x) &=& \bar{q}(x) \Gamma 
\frac{\lambda_k + i \lambda_l}{2} q(x), \nonumber \\
\Gamma &=& \left\{
\begin{array}{cc} 
\gamma_\mu & X = V_\mu \\
\gamma_\mu \gamma_5 & X = A_\mu \\
2 \gamma_5 & X = P\;. \\
\end{array}
\right.
\end{eqnarray}
\end{itemize}
From this list of $S$-matrix elements we have a redundant set of 19
equations for 14 coupling constants. Before presenting the results a few
comments are at hand, including some points we have mentioned already
before:
\begin{enumerate}
\item All one-loop diagrams with resonances have been expanded in powers of
$1/M^2_R$ with the heavy mass expansion. See Appendix \ref{appendixmass}
for details and examples. In particular, as we have verified, this
expansion ensures that the non-local contributions (e.g. to the scattering
amplitudes) reduce at order $O(e^2 p^2)$ to integrals which are independent
of the momentum transfer.
\item In order to renormalize the couplings $\hat K_i$ we work in $d$
dimensions and use the usual modified $\overline{MS}$ scheme, see
Eqs.(\ref{example1} - \ref{ms}). The set of matrix elements listed above
leads to a over-determined system of equations for the $K_i$. We have
checked that the renormalization of $\hat K_i$ is consistent in all
processes considered. Since we are only interested in the resonant
contributions we will not display them here.
\item The gauge fixing parameter $\lambda$ in the Lagrangian ${\cal
L}^\gamma_{kin}$ is set to $\lambda = 1$ (Feynman gauge). The matching
procedure with $S$-matrix elements ensures that we deal with gauge
independent quantities. The coupling constants $K_i$ themselves may be
gauge dependent, since in general they are not observables.
\item We have checked several parts of our calculation with results given
in the literature, namely $(i)$ the resonance-photon loops to
$M_{\pi^\pm}^2$ and $M_{K^\pm}^2$ from \cite{baur96} restricted to the
terms at order $O(e^2m_q)$, $(ii)$ the tree level contributions to all the
masses from \cite{urech95,neufeld95}, $(iii)$ the tree level calculation
for $F_{\pi^\pm}$ and the $K_{\ell 3}$ form factor $f_+^{K^0 \pi^-}(0)$ at
zero momentum transfer from \cite{neufeld95a}.
\item In Figure \ref{res:figmatch} we give as an example the one-loop
contributions to the scattering amplitude $\pi^- \pi^0 \to K^-
K^0$. Although the Lagrangian ${\cal L}^R_2$ that describes the interaction
between Goldstone bosons, photons and resonances is linear in the resonance
fields, in some graphs there occur vertices that couple two resonances to
Goldstone boson fields. These couplings (with a definite strength) are
generated by the covariant derivative (\ref{rkin}) acting on the
resonances. These diagrams are necessary in order to obtain a consistent
determination of the resonance contributions to the $K_i$. We have
restricted ourselves to the linear resonance model as defined in
(\ref{lr}). In particular we have not considered any extension to models
with multi-resonance couplings.
\item  Since we are only concerned with electromagnetic corrections and
work at order $O(e^2 p^2)$, we have neglected all effects due to the
mass difference $m_u - m_d$, i.e. contributions of the order
$O\left[e^2(m_u - m_d)\right]$ are assumed to be negligible.
Throughout we have worked in the isospin limit $m_u = m_d =\hat{m}$.
\item We would like to emphasize again that the pseudoscalar resonances do 
not contribute at the order $O(e^2p^2)$, thus we conclude that 
$K^{P}_{i}=0\; (i=1 \ldots 14)$.
\item In a calculation to the order $O(e^2 m_q)$ in CHPT with
electromagnetic interactions (but without resonances) we may distinguish
two kinds of terms: Loop contributions and counterterms (including the
$K_i$). For the matching to the resonant contributions only the terms
proportional to the $K_i$ are necessary and only those have been
calculated. However, one of us has checked that the loop contributions to
the electromagnetic mass shifts at order $O(e^2 m_q)$ are generated by the
resonances within photon loops at the two loop level \cite{baur96a}.
\end{enumerate}

\begin{table}[p]
\vspace{-10mm}
$$
\begin{array}{|l|r|r|r|r|r|}
\hline
   & \multicolumn{2}{|c|}{}   & \multicolumn{2}{|c|}{} & \\[-2mm]
\mbox{ Type: }   & \multicolumn{2}{|c|}{V+A}   & 
                   \multicolumn{2}{|c|}{S+S_1} & \mbox{ Total } 
\\[2mm] \hline &&&&&\\[-2mm]
\mbox{ Units: }   & \times 1/(4\pi)^2 & \times 10^{-3} & 
                    \times 1/(4\pi)^2 & \times 10^{-3} & \times 10^{-3} 
\\[2mm]
\hline &&&&&\\[-3mm]
K^R_1 (\mu_0)   & \displaystyle  -\frac{3}{16}\left( 5+\ln 2\right) 
                & - 6.8   
                & \displaystyle{\frac{3}{4} \frac{c_d^2}{F_0^2} 
                  \left( \ln \frac{M_S^2}{\mu_0^2}  +\frac{1}{6}  \right)}  
                & 0.4     
                & - 6.4      
\\[2mm] \hline  &&&&&\\[-3mm]   
K^R_2(\mu_0)    & \displaystyle -\frac{3}{16}\left( 3-\ln 2\right) 
                & - 2.7    
                & -\displaystyle{\frac{3}{4} \frac{c_d^2}{F_0^2} 
                  \left( \ln \frac{M_S^2}{\mu_0^2}  +\frac{1}{6}  \right)}  
                & - 0.4            
                & - 3.1
\\[2mm] \hline  &&&&&\\[-3mm]
K^R_3(\mu_0)    & \displaystyle  \frac{3}{16}\left( 5 + \ln 2 \right) 
                & 6.8      
                & \displaystyle{- \frac{3}{4} \frac{c_d^2}{F_0^2} 
                  \left( \ln \frac{M_S^2}{\mu_0^2}  +\frac{1}{6}  \right)}
                & - 0.4        
                & 6.4
\\[2mm] \hline  &&&&&\\[-3mm]
 K^R_4(\mu_0)   & \displaystyle -\frac{3}{8}\left( 3-\ln 2\right) 
                & - 5.5 
                & \displaystyle{- \frac{3}{2} \frac{c_d^2}{F_0^2} 
                  \left( \ln \frac{M_S^2}{\mu_0^2}  +\frac{1}{6}  \right)}   
                & - 0.7    
                & - 6.2
\\[2mm] \hline   &&&&&\\[-3mm]
K^R_5(\mu_0)    & \displaystyle  \frac{9}{16}\left( 5+\ln 2\right) 
                & 20.3   
                & \displaystyle{-\frac{3}{4} \frac{c_d^2}{F_0^2} 
                  \left( \ln \frac{M_S^2}{\mu_0^2}  +\frac{1}{6}  \right)}  
                & - 0.4     
                & 19.9
\\[2mm] \hline  &&&&&\\[-3mm]
K^R_6(\mu_0)    &\displaystyle \frac{9}{16} (3 - \ln 2) 
                & 8.2      
                & \displaystyle \frac{3}{4} \frac{c_d^2}{F_0^2} 
                  \left( \ln \frac{M_S^2}{\mu_0^2} + \frac{1}{6} \right)
                & 0.4          
                & 8.6   
\\[2mm] \hline  &&&&&\\[-3mm]
K^R_7 \dots K^R_{10} 
                & 0  & 0 & 0  & 0  & 0
\\[2mm] \hline  &&&&&\\[-3mm]
K^R_{11}(\mu_0) & \displaystyle \frac{3}{16} \left( \ln 
                  \frac{M_V^2}{\mu_0^2} + \frac{7}{6}  - \ln 2 \right)   
                & 0.6    & 0  & 0  & 0.6  
\\[2mm] \hline  &&&&&\\[-3mm]
K^R_{12}(\mu_0) & \displaystyle -\frac{3}{8} \left( \ln 
                  \frac{M_V^2}{\mu_0^2} + \frac{19}{6}  + \ln 2 \right)   
                & - 9.2  & 0  & 0  & - 9.2 
\\[2mm] \hline  &&&&&\\[-3mm]
K^R_{13}        & \displaystyle{\frac{9}{4}} 
                & 14.2   & 0  & 0  & 14.2 
 \\[3mm] \hline
\end{array}
$$
\caption[]{\label{silist} Contributions from vector and axial vector
resonances ($V+A$), and scalar octet and scalar singlet resonances
($S+S_1$) to the couplings $K_1 \dots K_{13}$. In the second and in the
fourth row we list the algebraic expressions, and in the third and in the
fifth row we indicate the corresponding numerical values at the scale point
$\mu_0 = M_V \simeq M_\rho$. In the last row we give the sum of the
numerical values.  We have used the relations (\ref{sumrules}) and the
numerical values for the parameters given in
(\ref{values},\ref{scalarvalues}).}
\end{table}

The full algebraic expressions of the $K^R_i$ (and for $C$) are given in
Appendix \ref{appendixkr}. In Table \ref{silist} we have listed the values
of the $K^R_i$, which were obtained by using the relations and masses shown
in (\ref{sumrules}) and (\ref{values}). The coupling $K^{R}_{14}$ has been
omitted, since it does not contribute to physical amplitudes.

At first sight we have found that the resonances do not contribute to
$K^R_7 \ldots K^{R}_{10}$ at all. Furthermore we find that the vector and
axial vector parts, except for $K^{R}_{11}$, are of the same order or even
larger as the value $1/(4\pi)^2 \simeq 6.3 \times 10^{-3}$ implied by naive
dimensional analysis \cite{manohar84} and dominate the scalar resonant
contributions, which are of the order $0.4\times 10^{-3}$. In most cases
the nonvanishing contributions are scale dependent. The scale dependence
of $K^R_1 \ldots K^R_6$ is due to the scalar resonances and that of
$K^{R}_{11}$ and $K^{R}_{12}$ is due to the the vector and axial vector
contributions. Varying the scale between the values $M_{\eta}$ and $1\mbox{
GeV}$ causes a change in the central values of $K^R_1 \ldots K^R_6$ of
roughly $10\%$. This is due to the smallness of the scalar
contributions. The same variation of $\mu_0$ causes a change in
$K^{R}_{12}$ of about $20\%$. Only the coupling $K^{R}_{11}$ is very
sensitive to the scale point $\mu_0$. Moving the scale point between the
bounds given above invokes a shift in $K^{R}_{11}$ of the same size as the
coupling itself. However, since there is a strong cancellation in
$K^{R}_{11}$ between the vector and axial vector resonance contributions,
this coupling is small compared to the other coupling constants.

From Table \ref{silist} we find that the following linear combinations are
scale independent
\begin{eqnarray}\label{sitotal}
S^R_1 = &K^R_1 + K^R_2&=   -\frac{3}{2}\frac{1}{16\pi^2},  \nonumber\\   
&&\nonumber\\     
S^R_2 = &K^R_5 + K^R_6&=   \frac{9}{2}\frac{1}{16\pi^2}, \\ 
&&\nonumber\\
S^R_3 =& - 2 K^R_3 + K^R_4&=-\frac{3}{16\pi^2}, \nonumber    
\end{eqnarray}
where $S_1 \ldots S_3$ have been originally defined in \cite{neufeld95}. It
is interesting to note that the $\mu_0$ independent contributions are pure
numbers, an effect of the relations (\ref{sumrules}) that we have used. The
effects of the values we obtained for the $K^R_i$ are shown in \mbox{Section
\ref{res:section5}}.

The behaviour of the $K_i$ in the large $N_C$ limit \cite{hooft74} is
derived in Appendix \ref{appendixnc}. We have found that the combinations
$K_1 + K_3$ and $2 K_2 - K_4$ and the coupling constants $K_7$ and $K_8$
are suppressed by $1/N_C$ with respect to all the other coupling constants
(including $K_1 \ldots K_4$). Indeed as one can see from Table \ref{silist}
the contributions from the resonances vanish in these particular
cases. Thus the correct large $N_C$ behaviour is a possible hint to
resonance saturation in the electromagnetic sector. However, the coupling
$K_8$ is part of the corrections of order $O(e^2 m_s)$ to the charged pion
mass. In \cite{urech95} an estimate of the coupling $K^r_8(\mu)$ was
given. There we neglected in the difference $M_{\pi^\pm}^2 - M_{\pi^0}^2$
the unknown counterterms proportional to $e^2 M_\pi^2$ and fixed the mass
difference at its experimental value. We found
\begin{equation}\label{k8}
K^r_8(M_\rho) = (- 4.0 \pm 1.7 ) \times 10^{-3}\quad .
\end{equation}
which is in contrast to our approach, where the coupling $K_8$ is not
affected by the resonances and thus $K^R_8 = 0$. There are two possible
answers (that do not exclude each other), namely
\begin{enumerate}
\item[$(i)$] the scale point has not been chosen appropriately. Moving to
$\mu = M_\eta$ we have
\begin{equation}
K^r_8(M_\eta) = (- 1.4 \pm 1.7 ) \times 10^{-3}\quad ,
\end{equation}    
which is compatible with zero within the error bars.
\item[$(ii)$] There are other contributions that generate the value given
in (\ref{k8}), which would show up in the remainder $\hat K_8$. If we refer
to the resonant contributions as long distance physics, possible additional
contributions could have their origin in the short distance range. Bijnens
evaluated the corrections to Dashen's theorem at order $O(e^2 m_s)$ within
the framework of the $1/N_C$ approach \cite{bijnens93}. He
performed explicitly a short distance calculation of the masses, where no
part proportional to $e^2 m_s$ has been found in the contributions to the
pion masses. This possibility has therefore no answer yet.
\end{enumerate}

We work at the order $O(e^2p^2)$ within the resonance approach determined
by the Lagrangians in (\ref{lr},\ref{rkin}). Higher order corrections arise
from two sources. $(i)$ The masses and the decay constants of the
resonances do not live in a $SU(3)$ invariant world. We know only very
little on these corrections, especially for the decay constants. For the
contributions to the masses of the vector resonances in the quark mass
expansion see \cite{gasser82,jenkins95}. $(ii)$ Higher order terms in the
heavy mass expansion may give considerably large corrections to the leading
order contribution, in particular if the expansion parameter is
$M_K^2/M_V^2$, see {\it Example 2} in Appendix \ref{appendixmass}. Thus,
when calculating processes with the numbers for the $K^R_i$ presented
above, one should keep in mind the possibility of higher order corrections,
especially when kaons are involved.

\section{Applications} 
\label{res:section5}

In this section we briefly give  the implications of our results on
the $K^R_i$ to several matrix elements and ratios thereof quoted in the
literature. We will display the counterterms at order $O(e^2 p^2)$
proportional to $K_i$ only, for the full expressions we refer to the
indicated literature.

There are three combinations of form factors and decay constants that
involve the same counterterms \cite{neufeld95,neufeld95a},
\begin{eqnarray}
r_{K \pi}|_{Res.} & = & \left. 
                        \frac{f_+^{K^+ \pi^0}(0)}{f_+^{K^0 \pi^-}(0)}
                        \right|_{Res.}\; = \;
                        \frac{4}{9} e^2 \frac{M_K^2}{M_\eta^2 - M_\pi^2}
                        \left( 2 S^R_2 + 3 S^R_3 \right) \nonumber \\
\left.f_+^{\eta \pi^+}(0)\right|_{Res.}& = & 
                         \frac{1}{\sqrt{3}} r_{K \pi}|_{Res.} \nonumber \\
\left.\frac{F_{K^0}F_{\pi^\pm}}{F_{K^\pm}F_{\pi^0}(\lambda_3/\sqrt{2})}
   \right|_{Res.} & = & - \frac{1}{3} e^2 
                        \left( 2 S^R_2 + 3 S^R_3 \right)\;,
\end{eqnarray}
where $f^{P \pi}_+(0)$ is the form factor associated with the momentum
$(p_P + p_\pi)_\mu$ in the decay $P \to \pi \ell \overline{\nu}_\ell$ at
zero momentum transfer and $F_P$ is the $P_{\ell 2}$ decay constant. From
Eq.(\ref{sitotal}) we find immediately that $2 S^R_2 + 3 S^R_3 = 0$,
therefore the above quantities do not receive contributions from the
resonances. Next we consider the resonant part of the electromagnetic
corrections to the amplitude $\eta \to 3 \pi$ \cite{baur96b},
\begin{eqnarray}
A_{QED}(s,t,u)|_{Res.} & = & - \frac{2}{9\sqrt{3}} \frac{e^2}{F_0^2}
               M_\pi^2 \left[ 1 + \frac{3(s - M_\pi^2) 
               - M_\eta^2}{M_\eta^2 - M_\pi^2}\right] \nonumber \\
&& \times \left( 2 S^R_2 + 3 S^R_3 - 2 K^R_9 - 2 K^R_{10} \right)\;,
\end{eqnarray}
which again vanish in the resonance approach. 

Finally, we discuss the resonant contributions to the masses. We start with
the masses of the neutral particles \cite{urech95},
\begin{eqnarray}
M_{\pi^0}^2|_{Res.} & = & - \frac{2}{9} e^2 M_\pi^2 \left( 12 S^R_1 
   + 10 S^R_2 + 9 S^R_3 - 12 K^R_7 - 12 K^R_8 - 10 K^R_9 
   - 10 K^R_{10} \right) 
\nonumber \\
M_{K^0}^2|_{Res.} & = & - \frac{8}{9} e^2 M_K^2 \left( 3 S^R_1 + S^R_2 
   - 3 K^R_7 - 3 K^R_8 - K^R_9 - K^R_{10} \right) 
\nonumber \\
M_\eta^2|_{Res.} & = & \frac{2}{9} e^2 \left[ 2 M_\pi^2 \left( 
   K^R_9 + K^R_{10} \right) - M_\eta^2 \left( 12 S^R_1 + 6 S^R_2 
   + 3 S^R_3 - 12 K^R_7 \right. \right. \nonumber \\
&&\left. \left. \hspace{5mm} 
   - 12 K^R_8 - 4 K^R_9 - 4 K^R_{10} \right) \right]\;.
\end{eqnarray}
As the quantities in the previous examples, the masses of the neutral
particles are unaffected by the resonances, whereas the masses of the
charged particles have the form,
\begin{eqnarray} 
M_{\pi^\pm}^2|_{Res.}(\mu_0) & = & 4 e^2 \left\{ 2 M_K^2 K^R_8
   - \frac{1}{9} M_\pi^2 \left[ 6 S^R_1 + 5 S^R_2 - 6 K^R_7 - 15 K^R_8 
   \right. \right. \nonumber \\
&& \left. \left. \hspace{1cm}
   - 5 K^R_9 - 23 K^R_{10} - 18 K^R_{11}(\mu_0) \right]  \right\} 
\nonumber \\
M_{K^\pm}^2|_{Res.}(\mu_0) & = & \frac{4}{3} e^2 \left\{M_\pi^2 
   \left( 3 K^R_8 + K^R_9 + K^R_{10} \right) \right. \nonumber \\ 
&& \left. \hspace{1cm}   
- \frac{1}{3} M_K^2 \left[ 6 S^R_1 + 5 S^R_2 - 6 K^R_7 - 24 K^R_8 
\right. \right. \nonumber \\
&& \left. \left. \hspace{1cm} - 2 K^R_9 - 20 K^R_{10} - 18 K^R_{11}(\mu_0) 
   \right] \right\}\;. 
\end{eqnarray}
They receive contributions from the resonances and all the scale dependence
is concentrated in $K^R_{11}(\mu_0)$. Putting in numbers, we observe that
the corrections to $M_{\pi^\pm}^2$ and $M_{K^\pm}^2$ are proportional to
$M_\pi^2$ and $M_K^2$, respectively, and that they contain both the same
coefficient,
\begin{eqnarray}\label{ratio}
\left.\frac{M_{\pi^\pm}^2}{M_\pi^2} \right|_{Res.}(\mu_0) 
& = & \left.\frac{M_{K^\pm}^2}{M_K^2} \right|_{Res.}(\mu_0) \nonumber \\
& = & - \frac{1}{64\pi^2} e^2 \left[ 17 + 6 
      \left( \ln 2 - \ln \frac{M^2_V}{\mu_0^2} \right) \right] \quad .
\end{eqnarray}

Now we are able to discuss the different terms that contribute to the
corrections to Dashen's theorem \cite{dashen69,leutwyler95}, that involves
the electromagnetic mass differences $(M_{K^\pm}^2 - M_{K^0}^2)_{e.m.} -
(M_{\pi^\pm}^2 - M_{\pi^0}^2)_{e.m.} = (\Delta M_K^2 - \Delta
M_\pi^2)_{e.m.}$ at order $O(e^2m_q)$. We do not show the algebraic
expression here, in its full form it may be found in \cite{urech95}. We
distinguish two different parts, one is generated by the photon loop and
the counterterms $K_i$, the other part is due to mesonic loops and to the
counterterm $L_5$ and therefore proportional to the coupling $C$. The two
parts are then again separated in contributions from the loops and from the
counterterms, respectively. As an ingredient we need the value of
$L_5^r(\mu)$ \cite{gasser84},
\begin{equation}
L_5^r(\mu) = (1.4 \pm 0.5) \times 10^{-3} - \frac{3}{128\pi^2} 
             \ln \frac{\mu}{M_\rho}\;.
\end{equation}

\begin{table}[t]
$$
\begin{array}{|c|r|r|r|r|r|}
\hline
\multicolumn{6}{|c|}{}\\[-2mm]
\multicolumn{6}{|c|}{(\Delta M_K^2 - \Delta M_\pi^2)_{e.m.} \quad
                     \left[ \times 10^{-3}\; ({\rm GeV})^2 \right]}
\\[2mm] \hline 
   & \multicolumn{2}{|c|}{}   
   & \multicolumn{2}{|c|}{} 
   &  
\\[-2mm]
   & \multicolumn{2}{|c|}{\sim e^2 m_q }
   & \multicolumn{2}{|c|}{\displaystyle \sim e^2 \frac{C}{F_0^4} m_q } 
   &  
\\[4mm] \cline{2-3} \cline{3-5} 
&&&&&\\[-2mm]
\mu  & \multicolumn{1}{|c|}{\mbox{ Loops }} 
     & \multicolumn{1}{|c|}{\Delta R_{Res.}}
     & \multicolumn{1}{|c|}{\mbox{ Loops }} 
     & \multicolumn{1}{|c|}{L_5^r}
     & \multicolumn{1}{|c|}{\mbox{ Total }}
\\[2mm] \hline &&&&&\\[-2mm]
M_\eta       & 0.52  & - 0.70 & - 0.04 & - 0.67  & - 0.88             
\\[2mm] \hline &&&&&\\[-2mm]   
M_\rho       & 0.79  & - 0.70 &   0.11 & - 0.43  & - 0.21 
\\[2mm] \hline &&&&&\\[-2mm]
1\;{\rm GeV} & 1.0   & - 0.70 &   0.23 & - 0.24  &  0.30
\\[2mm] \hline &&&&&\\[-2mm]
{\mbox{ Errors }} &  & {}^{\;+ 0.13}_{\;- 0.10}  &  & \pm 0.15 
                  & {}^{\;+ 0.28}_{\;- 0.25} 
\\[2mm] \hline
\end{array}
$$
\caption[]{\label{dashen}
Corrections to Dashen's theorem at order $O(e^2 m_q)$ at
different scale points $\mu$. The contributions from the resonances $\Delta
R_{Res.}$ are $\mu$ - independent and renormalized at $\mu_0 = M_\rho$. The
error of this quantity reflects the variation of the renormalization scale
$\mu_0$ from $M_\eta\;(+)$ up to $1\;{\rm GeV}\;(-)$. The error of $L_5^r$
is $\mu$ - independent. }
\end{table}

The numerical results are presented in Table \ref{dashen} for the three
scale points $\mu = (M_\eta, M_\rho, 1\;{\rm GeV})$. The contributions from
the resonances, shown as counterterms $\Delta R_{Res.}$, depend on the
renormalization scale $\mu_0$ that we choose to be $\mu_0 = M_\rho$. The
variation of this scale is reflected in the error of this value, namely the
upper bound corresponds to $\mu_0 = M_\eta$, the lower bound to $\mu_0 =
1\;{\rm GeV}$, respectively. We obtain
\begin{equation}\label{deltar}
\Delta R_{Res.} = \left(- 0.70^{\;+ 0.13}_{\;- 0.10} \right)
                  \times 10^{-3}\; ({\rm GeV})^2 \;.
\end{equation}
From the results in Table \ref{dashen} we see that all the possible
contributions are of the same magnitude, none of them is
negligible. Considering a scale point in the range $0.7 \sim 1 \;{\rm
GeV}$, the uncertainty that enters the calculation at the order $O(e^2
m_q)$ is of the same size as the central value itself.

In one place we are able to check the size of higher order corrections. In
\cite{baur96} (see also \cite{donoghue93}) we have calculated with the
Lagrangians in (\ref{lr},\ref{rkin}) the contributions from the resonances
to electromagnetic mass differences without truncating the integrals at the
order $O(e^2 m_q)$. At the scale point $\mu = \mu_0 = M_\rho$ we have found
for the corrections to Dashen's theorem (transformed to the present values
for the parameters)
\begin{equation}
\left( \Delta M_K^2 - \Delta M_\pi^2 \right)_{e.m.}(M_\rho) = 0.37 
      \times 10^{-3}\; ({\rm GeV})^2\hspace{1cm}\mbox{in \cite{baur96}}.
\end{equation} 
Note that this number is very sensitive to the choice of the parameters in
the resonance sector. In Table \ref{dashen} the corresponding result is the
sum of the contributions to the photon loop and $\Delta R_{Res.}$, namely
\begin{equation}
\left.\left( \Delta M_K^2 - \Delta M_\pi^2 \right) 
      \right|_{\gamma - loop\, +\, Res.}(M_\rho) = 
0.09 \times 10^{-3}\; ({\rm GeV})^2 
\end{equation}
at the same scale and renormalization point as above. Thus the higher order
terms from the exact evaluation of the integral gives $40\%$ corrections to
$\Delta R_{Res.}$ as it could be expected from {\it Example 2} in 
Appendix \ref{appendixmass}. Due to the strong cancellation in the sum of
the loop and counterterm contributions, it completely reveals the number of
this part and moves as a consequence the central value of the final result to
the positive region. Taking this higher order corrections explicitly into
account we arrive at
\begin{eqnarray}
\lefteqn{\left( \Delta M_K^2 - \Delta M_\pi^2 \right)_{e.m.}(M_\rho) =}
\nonumber \\[2mm]  
&&\left( 0.07^{\;+ 0.28}_{\;- 0.25} \right) \times 10^{-3}\; ({\rm GeV})^2 
+ \Delta R_{Non-res.}(M_\rho) + O(e^2 m_q^2)\;,
\end{eqnarray}
where $\Delta R_{Non-res.}$ are the non-resonant contributions at the order
$O(e^2 m_q)$ determined by the $\hat{K}_i(\mu,\mu_0)$ defined in
(\ref{ansatz2}) that we do not consider here. Their determination is beyond
the scope of this work.

In summary we find that all the contributions to the corrections to
Dashen's theorem that we take into account are of the same
size. Furthermore we find that there occur strong cancellations in summing
up all the parts, leading to the result that the uncertainties at the order
$O(e^2 m_q)$ and higher order corrections could easily reveal the central
values quoted in Table \ref{dashen}; the same may be valid for non-resonant
contributions $\Delta R_{Non-res.}$\,. Thus from our calculation we may
conclude that the low energy contributions to the violation of Dashen's
theorem are small.

\section{Conclusions} 
\label{res:section6}

\hspace{6mm}1.\, Within the framework of chiral perturbation theory we have
considered the coupling constants $K_i\; (i = 1 \ldots 14)$ in the
electromagnetic interaction Lagrangian ${\cal L}^Q _4$ at order $O(e^2
p^2)$ \cite{urech95,neufeld95,baur96b}. We have determined the
contributions to the $K_i$ that arise from resonances within a photon loop
at order $O(e^2 p^2)$ by calculating masses, scattering amplitudes and
matrix elements with external currents (see Section
\ref{res:section4}). For the interactions of the pseudoscalar mesons with
the resonances we have used the lowest order interaction Lagrangian that is
linear in the resonance fields, introduced by Ecker {\it et al.}
\cite{ecker89}. Throughout we have worked in the isospin limit $m_u = m_d =
\hat{m}$ (i.e. neglecting contributions of the order $O\left[e^2(m_u -
m_d)\right]$).

2.\, Formally, we have split the contributions to the $K_i$ in a resonant
part $K^R_i$, which we have calculated, and a remainder $\hat K_i$. In
general the resonance-photon loops generate ultraviolet divergences that
are absorbed by renormalization of the corresponding $\hat K_i$ at a
specific scale point $\mu_0$. There are no contributions to $K_7 \ldots
K_{10}$ from the resonances at all (see Table \ref{silist}). For the
remaining $K^R_i$ we have found that the vector and axial vector parts,
except for $K^{R}_{11}$, are of the same order or even larger as the value
$1/(4\pi)^2 \simeq 6.3 \times 10^{-3}$ implied by naive dimensional
analysis \cite{manohar84}. Whenever the scalars contribute, they generate
small but scale dependent shifts.  Since the scalar contributions are
small, the values of $K^{R}_{1} \ldots K^{R}_{6}$ are hardly affected by
the scale dependence. This is in contrast to the case of $K^{R}_{11}$. Here
the scale dependence is due to the vector and axial vector
contributions. The constant $K^{R}_{11}$ is very sensitive to the choice of
the scale point $\mu_0$. Varying $\mu_0$ between the values $M_\eta$ and
$1\;{\rm GeV}$ gives rise to a shift in $K^{R}_{11}$ of the same size as
the coupling itself. However, since there is a strong cancellation in
$K^{R}_{11}$ between the vector and axial vector resonance contributions,
this coupling is small compared to the other coupling constants. The scale
dependence in the coupling $K^{R}_{12}$ is also generated by the vector and
axial vector contributions and the variation of $\mu_0$ between the bounds
given above causes a shift of about $20\%$.

3.\, Comparing our results to the expectation from the large $N_C$ limit
(see Appendix \ref{appendixnc}) we find agreement for the four linear
combinations of couplings that are suppressed in this limit. The couplings
$K_7$ and $K_8$ do not get any contributions from the resonances at
all. The combinations $K_1^R + K_3^R$ and $2K^{R}_{2}-K^{R}_{4}$ vanish,
even though the individual couplings $K_1 \ldots K_4$ each get large
contributions from the resonances.

4.\, In the evaluation of the resonance contributions to the $K_i$ we have
used the heavy mass expansion (see Appendix \ref{appendixmass}) in order to
truncate the integrals at the order $O(e^2 p^2)$. We have found that higher
order corrections can be sizeable, especially when the ratio $M_K^2/M_V^2$
is involved as expansion parameter, see {\it Example 2} in Appendix
\ref{appendixmass}. Thus when calculating processes with the values for the
$K^R_i$ presented in this article, one should keep in mind the possibility
of higher order corrections, in particular when kaons are involved.

5.\, As an application we have discussed the effects of our result on the
corrections to Dashen's theorem \cite{dashen69,leutwyler95} at the order
$O(e^2 m_q)$.  We find a strong cancellation in the sum of the different
terms leading to the result that the uncertainties are of the same order of
magnitude than the central value at the scale point $\mu = M_\rho$ (the
same holds for all the scale points in the range $0.7 \sim 1\;{\rm
GeV}$),
\begin{eqnarray}\label{dashencorr}
\lefteqn{\left( \Delta M_K^2 - \Delta M_\pi^2 \right)_{e.m.}(M_\rho) =}
\nonumber \\[2mm]  
&&\left( 0.07^{\;+ 0.28}_{\;- 0.25} \right) \times 10^{-3}\; ({\rm GeV})^2 
+ \Delta R_{Non-res.}(M_\rho) + O(e^2 m_q^2)\;,
\end{eqnarray}
where $\Delta R_{Non-res.}$ are the non-resonant contributions at the order
$O(e^2 m_q)$ determined by the $\hat{K}_i$. Their inclusion is beyond the
scope of this work. The error indicated is due to the strong coupling
constant $L_5^r(M_\rho)$ and the variation of the renormalization scale
$\mu_0$ for the resonance contributions in the range $M_\eta \sim 1\;{\rm
GeV}$. Note that in the numerical part in (\ref{dashencorr}) the higher order
corrections due to the heavy mass expansion are already included
\cite{baur96}.\\[2mm]

After completing this article we became aware of a work by Bijnens and
Prades \cite{bijnens96} on the same subject. The authors have considered
the electromagnetic interaction Lagrangian ${\cal L}_4^Q$ including the
$\eta'$ as an additional degree of freedom. Within the $1/N_C$ approach
\cite{bardeen89,bijnens93,bardeen87} they have calculated the short
distance contributions to all the coupling constants $K_i$ at leading order
in $1/N_C$ and the long distance part of the linear combinations of $K_i$
needed for the electromagnetic mass corrections. They found a large
violation of Dashen's theorem,
\begin{equation}
\left( \Delta M_K^2 - \Delta M_\pi^2 \right)_{e.m.} 
=  ( 1.06 \pm 0.32 ) \times 10^{-3}\; ({\rm GeV})^2\;.
\end{equation}
The source of the large difference to our result may be identified
easily. The contributions from the $K_i$ (that we call $\Delta R$) is
positive in their approach \cite{bijnens96},
\begin{equation}
\Delta R  =  0.53 \times 10^{-3}\; ({\rm GeV})^2\;,
\end{equation}
whereas the corresponding value of our calculation is given in
Eq.(\ref{deltar}), $\Delta R_{Res.} = - 0.70 \times 10^{-3}\; ({\rm
GeV})^2$. Since their technique involves an Euclidean cut-off, the
comparison to our procedure cannot be carried out directly. It will need
some more work to find the connection between the two approaches.

Even more recently two other preprints by Gao, Yan and Li \cite{gao96} and
by Donoghue and P\'{e}rez \cite{donoghue96} concerning the electromagnetic
mass differences came to our attention. In \cite{gao96} the authors
consider an $U(3)_R \times U(3)_L$ effective field theory containing the
low-lying pseudoscalar, vector and axial vector mesons. The divergences
that occur in the resonance-photon loop contributions to the
electromagnetic mass corrections are absorbed by using an intrinsic
parameter $g$ of the theory \cite{gao96}. The corrections to Dashen's
theorem are found to be large,
\begin{equation}
\left( \Delta M_K^2 - \Delta M_\pi^2 \right)_{e.m.} 
 =  1.08 \times 10^{-3}\; ({\rm GeV})^2 \;.
\end{equation}
In \cite{donoghue96} the Cottingham method \cite{cottingham63} is used in
order to calculate the electromagnetic mass corrections. This approach
leads to a dispersion integral over the Compton scattering amplitudes
$\gamma (\pi,K) \to \gamma (\pi,K)$. As input serves CHPT for the elastic
scattering, inelastic production of the low-lying resonances and
experimental data in the intermediate energy region, and perturbative QCD
at high energies. The authors have found substantial contributions to the
electromagnetic kaon mass difference mainly due to the effect of the kaon
mass in the propagator of the Born term in the $\gamma K$ scattering
amplitude \cite{donoghue96} (this corresponds to the contribution from
graph a) in Figure \ref{appd:fig1} in the present work). Correspondingly
the corrections to Dashen's theorem are also large,
\begin{equation}
\left( \Delta M_K^2 - \Delta M_\pi^2 \right)_{e.m.}
  =  ( 1.34 \pm 0.61 ) \times 10^{-3}\; ({\rm GeV})^2\;,
\end{equation}
where we combined the individual electromagnetic mass differences and added
the errors in quadrature. One difference to our work may be traced out in
the resonance region, where in \cite{donoghue96} the experimental masses
and decay constants have been applied (including finite widths), whereas we
work in the $SU(3)$ limit and neglect the widths of the resonances. Whether
this different treatment of the resonances is essential remains to be
studied.

\section*{Acknowledgments}

We are grateful to J. Bijnens, J. Gasser and D. Wyler for very helpful
discussions. R.U. would like to thank Matthias Steinhauser for his never
ending patience to explain the virtue of the heavy mass expansion.

\appendix
\section{Heavy Mass Expansion}
\label{appendixmass} 

With the heavy mass expansion loop integrals may be conveniently
simplified, in the case when all the internal masses can be divided into a
set of large \mbox{$\underline{M}=\{M_{1},M_{2}, \ldots$\}} and small
$\underline{m}=\{m_{1},m_{2}, \ldots\}$ , and when all external momenta
$\underline{q}=\{q_{1},q_{2}, \ldots\}$ are small compared to the scale of
the large masses $\underline M$. The heavy mass expansion was first
described in an equivalent form in \cite{gorishny89} and was mathematically
rigorously formulated in \cite{gorishny87,pivovarov93,smirnov90}. Let
$F_{\Gamma}$ be the Feynman integral associated with the Feynman diagram
$\Gamma$, then the proposition is valid \cite{smirnov95}
\begin{equation}
F_{\Gamma} \stackrel{\underline{M} \to \infty}{=} 
\sum_{\gamma} F_{\Gamma / \gamma} \circ
T_{\underline{q}^{\gamma},\underline{m}^{\gamma}} 
F_{\gamma}(\underline{q}^{\gamma},\underline{m}^{\gamma},\underline{M})
\end{equation}
where the sum is over all subgraphs $\gamma$ of $\Gamma$ such that each
$\gamma$
\begin{description}
\item[a)] contains all lines with heavy masses,
\item[b)] consists of connectivity components that are
one-particle-irreducible with respect to lines with small masses.
\end{description}
The operator $T$ performs Taylor expansion in the variables
$\underline{q}^{\gamma}$ and $\underline{m}^{\gamma}$. It is understood
that the operator $T$ acts on the integrand of the subgraph $\gamma$ thus
producing a polynomial $P_{\gamma}$ in $\underline{q}^{\gamma}$ and
$\underline{m}^{\gamma}$. The momenta $\underline{q}^{\gamma}$ denotes the
external momenta of the subgraph $\gamma$. It should be noted that the
external momenta $\underline{q}^{\gamma}$ is defined with respect to
$\gamma$ and thus in general consists of some genuine external momenta,
shared by the whole Feynman diagram $\Gamma$ and $\gamma$, as well as
momenta flowing through internal lines of $\Gamma$, which are external ones
of $\gamma$. The masses $\underline{m}^{\gamma}$ collects all light masses
in the subgraph $\gamma$. The notation $F_{\Gamma / \gamma}$ means that the
subgraph $\gamma$ is contracted to a single vertex $v_{\gamma}$. After
insertion of the polynomial $P_{\gamma}$ in $v_{\gamma}$ the new Feynman
integral $F_{\Gamma / \gamma} \circ P_{\gamma}$ must be evaluated. All
possible subgraphs $\gamma$ must be found and their contribution must be
added to the result. Writing the Taylor expansion operator as
\begin{equation}
T_{x_{1},x_{2}, \ldots} = \sum^{\infty}_{n=0} t^{n}(x_{1},x_{2}, \ldots)
\end{equation}
then one immediately sees that
\begin{eqnarray}
t^{n}(\underline{q}^{\gamma},\underline{m}^{\gamma})
F_{\gamma}(\underline{q}^{\gamma},\underline{m}^{\gamma},\underline M)
\nonumber
\end{eqnarray}
scales like $M^{d(\gamma)-n}$, where $d(\gamma)$ is the mass dimension of
the subgraph $\gamma$.
\begin{figure}[!t]
\begin{center}
\leavevmode
\epsfig{file=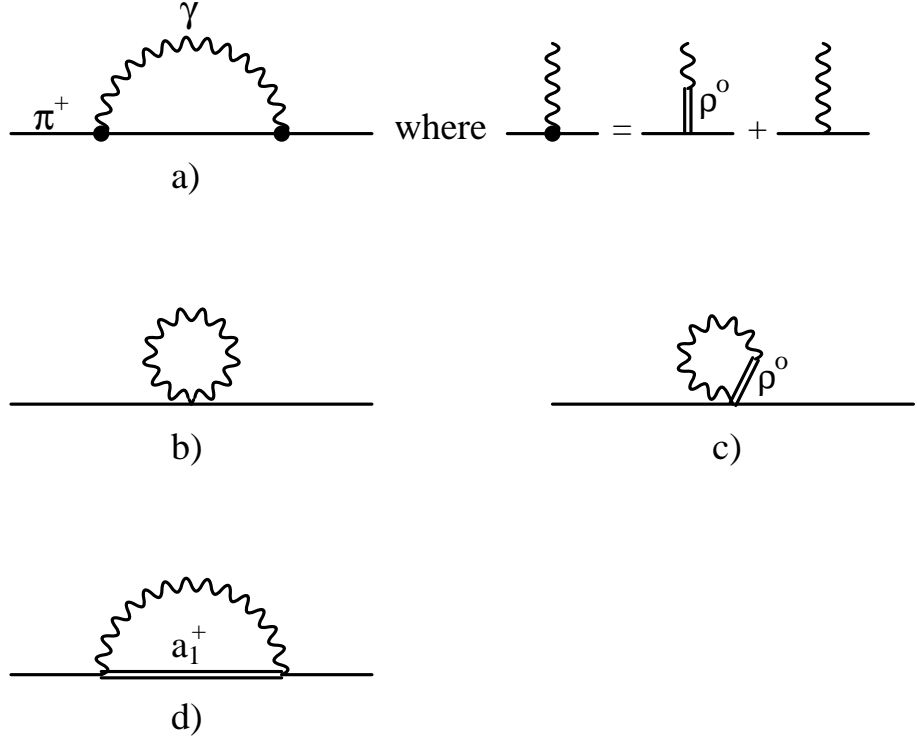,width=12cm}
\caption[]{\label{appd:fig1}
One-loop contributions to the
electromagnetic mass shift of $\pi^\pm$. }
\end{center}
\vspace{1cm}
\end{figure}

\begin{figure}[!t]
\begin{center}
\leavevmode
\epsfig{file=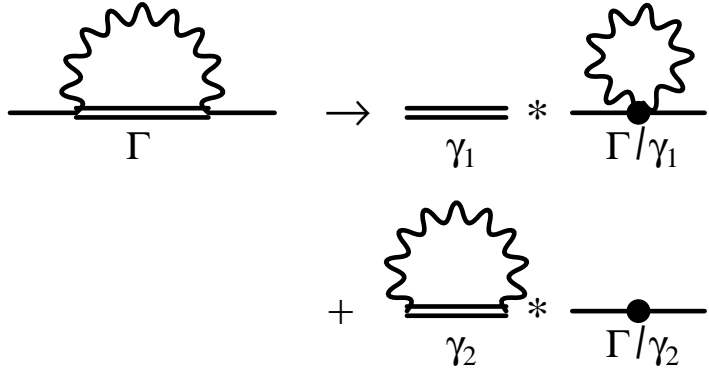,height=5cm}
\caption[]{\label{appd2:fig2}
Example of the heavy mass expansion.}
\end{center}
\vspace{1cm}
\end{figure}

{\it Example 1}: Consider the loop contribution to $M_{\pi^\pm}^2$
containing the axial vector resonance $a_1$ (with mass $M_A$) shown in
Figure \ref{appd:fig1}d). This diagram can be worked out exactly and gives
\cite{baur96}
\begin{eqnarray}\label{example1}
F_\Gamma & = &  (3-\epsilon) \frac{e^2 F^2_A}{F^2_0}\frac{1}{i} \int 
              \frac{d^4 q}{(2\pi)^4}\frac{1}{M^2_A-q^2} \nonumber \\
        && + \frac{e^2 F^2_A}{F^2_0}\frac{1}{i}
        \int \frac{d^4 q}{(2\pi)^4}\frac{q^2\left[M^2_\pi
                  +(3-\epsilon)\nu \right]
        +(2-\epsilon)\nu^2}{q^2\left[M^2_A-(q+p)^2\right]} \quad ,
\end{eqnarray}
where $\epsilon = 4-d$ ($d$: dimension), $\nu = pq$ and $p$ is the pion
momentum. The integrals are easily evaluated,
\begin{eqnarray}\label{appd:eq1}
F_\Gamma & = & \frac{3e^2}{F^2_0 16\pi^2}\left[ F^2_A 
            M^2_A\left(\ln\frac{M^2_A}{\mu_0^2}
            + \frac{2}{3}\right)\right] 
 +\frac{6e^2}{F^2_0}F^2_A M^2_A\lambda \nonumber \\
         && - \frac{e^2 F^2_A}{F^2_0 16\pi^2} M^2_\pi
           \left[2+\frac{3}{2}\ln\frac{M^2_A}{\mu_0^2}
           +I_1\left(\frac{M^2_\pi}{M^2_A}\right)\right] \quad ,       
\end{eqnarray}
with
\begin{eqnarray}\label{ms}
\lambda & = & \frac{\mu_0^{d-4}}{16\pi^2} \left\{\frac{1}{d-4}
                   -\frac{1}{2}[\ln 4\pi+\Gamma'(1) +1 ] \right\} 
\nonumber \\
I_1(z) & = & \int^1_0 x\ln[x-x(1-x)z] \,dx \quad .
\end{eqnarray}
In the heavy mass expansion two subgraphs $\gamma_{1}$ and$\gamma_{2}$ are
identified, see Figure \ref{appd2:fig2}.  The first subdiagram $\gamma_{1}$ is
found to be the heavy mass propagator. It is characterized by $q \ll M_A$.
The external momentum flowing through the heavy propagator is the loop
momentum $q$. We thus expand $\gamma_{1}$ in powers of $q$. Working up to
order $O(1/M_A^2)$ we get
\begin{equation}
P_{\gamma_{1}}= T_{k} F_{\gamma_{1}}=\frac{1}{M_A^2}+O(\frac{q^2}{M_A^2}) 
\quad .
\end{equation} 
This term corresponds to the naive expansion of a Feynman integral in
powers of $1/M_A^2$.  The diagram $\Gamma /\gamma_1$ vanishes,
however, in dimensional regularization, because only the massless photon
runs through the loop.  The second subgraph $\gamma_{2}$ is the loop
itself. $\gamma_{2}$ is characterized by $p \ll q,M_A$ and if the
photon had a mass $m_\gamma$ by $m_\gamma,p \ll q,M_A$. Expanding in
$\gamma_{2}$ the heavy propagator in $p$ to the order $O(1/M^{2}_{A})$ one
obtains
\begin{eqnarray}\label{appd:eq2} 
\sum_{\gamma} F_{\Gamma / \gamma} \circ 
T_{\underline{q}^{\gamma},\underline{m}^{\gamma}} & = &
F_{\Gamma / \gamma_{2}} T_{p} F_{\gamma_{2}} \nonumber \\
& = & \frac{3e^2}{F^2_0 16\pi^2}\left[ F^2_A M^2_A 
      \left(\ln\frac{M^2_A}{\mu_0^2} + \frac{2}{3}\right)\right] 
      + \frac{6e^2}{F^2_0}F^2_A M^2_A\lambda \nonumber \\
&& - \frac{e^2 F^2_A}{F^2_0 16\pi^2} M^2_\pi \left[2 
   + \frac{3}{2}\ln\frac{M^2_A}{\mu_0^2} - \frac{1}{4}\right]\;.
\end{eqnarray}
As can be seen from the equations (\ref{appd:eq1}) and (\ref{appd:eq2}) one
has to compare the numerical value of $I_1 (M^2_\pi/M^2_A)$ with $ -
1/4\,$. We may split the integral $I_1$ in a part at $z=0$ and a remainder
$\bar{I}_1 (z) = I_1 (z) - I_1 (0)$. Numerically we find
\begin{eqnarray}
I_1\left(\frac{M^2_\pi}{M^2_A}\right) & = & I_1 ( 0 ) 
      + \bar{I}_1 \left(\frac{M^2_\pi}{M^2_A}\right)\nonumber \\
& = & - 0.250 - 0.003\;,
\end{eqnarray} 
where we have used the relations in (\ref{values}). Thus in our case the
result from the heavy mass expansion in (\ref{appd:eq2}) corresponds to
$I_1(0)$.  The difference $\bar{I}_1 (z)$ to the exact result is
small. Indeed, inserting in (\ref{appd:eq1}) one gets for the expression in
the parentheses on the last line,
\begin{equation}
\left. \left[ 2 + \frac{3}{2}\ln\frac{M^2_A}{\mu_0^2} + I_1 (0) +
\bar{I}_1\left(\frac{M^2_\pi}{M^2_A}\right) \right]
\right|_{\mu_0 = M_\rho} = 2.790 - 0.003\;,
\end{equation}
where we have split the remainder from all the other contributions and
fixed the renormalization scale point at $\mu_0 = M_\rho$.  

{\it Example 2}: In the previous example we have considered an integral
that converges well in the $1/M_R$ expansion, i.e. the truncation after the
leading term gives a good description of the entire integral. In general,
we cannot expect that this approximation of a loop contribution works so
well. In the present example we give an estimate of the upper bound for the
error that we induce with the truncation. We consider again the loop
contributions in Figure \ref{appd:fig1}, but this time we choose the part a)
and calculate its contribution to $M_{K^\pm}^2$. The expansion parameter is
therefore $M_K^2/M_V^2 \sim 0.4$, and the relevant expression reads
\cite{baur96},
\begin{eqnarray}\label{appd:eq3} 
F_\Gamma & = & \frac{1}{8\pi^2} e^2 M^2_K \left[ \frac{7}{2} 
               - \frac{3}{2}\ln\frac{M^2_K}{M^2_V} 
               + I_2\left(\frac{M^2_K}{M^2_V}\right)\right] \nonumber \\ 
I_2(z) & = & \int^1_0 (1 + x) \left\{ ln[x + (1-x)^2 z] 
              - \frac{x}{x + (1 - x)^2 z}\right\} \,dx \; .  
\end{eqnarray} 
Note that $F_\Gamma$ is finite and therefore scale independent. Expanding
in $1/M_V^2$ leads in an analogous way as before to an expansion of the
integral $I_2(z)$ around $z=0$,
\begin{eqnarray}
I_2\left(\frac{M^2_K}{M^2_V}\right) & = & I_2(0) 
      + \bar{I}_2\left(\frac{M^2_K}{M^2_V}\right) \nonumber \\
& = & - 2.75 + 0.95\;,
\end{eqnarray}
where $\bar{I}_2(z)$ is defined similarly to $\bar{I}_1(z)$. The remainder
decreases the approximate result $I_2(0)$ by about $35\%$. Inserted in
(\ref{appd:eq3}) we find
\begin{equation}\label{para}
\left[ \frac{7}{2} - \frac{3}{2}\ln\frac{M^2_K}{M^2_V} + I_2(0)
+ \bar{I}_2\left(\frac{M^2_K}{M^2_V} \right) \right] = 2.08 + 0.95\;,
\end{equation}
where again we have given the remainder separately. The change from the
leading order term in the heavy mass expansion to the exact result is
large, about $46\%$.   
 
Therefore we conclude that higher order corrections to the leading order
term in the heavy mass expansion may be large, depending on the size of the
expansion parameter. The ratio $M_K^2/M_V^2$ is the largest value that
enters our calculations, leading to corrections of the order of $50\%$ in
the example above.    

\newpage
\section{Contributions to $K^R_i$}
\label{appendixkr}

Below we list  the finite contributions to the $K^R_i$ in their most general
form, i.e. without using the relations in Eq.(\ref{sumrules})
for the parameters in the resonance sector. For
completeness we give the contributions to the leading order constant $C$
\cite{ecker89}. Instead of $C$ we indicate the dimensionless quantity
$Z=C/F_0^4$. Vanishing $K^R_i$ are omitted from the list.
\vspace{3cm}

\underline{\em Vector Mesons}\hfill

\begin{eqnarray}
\begin{array}{ll}
\underline{F_V^2}: & \\[4mm] 
         \multicolumn{2}{l}{\displaystyle Z^V(\mu_0) = 
                      - \frac{3}{2}\frac{1}{16\pi^2}\frac{F_V^2}{F_0^4}
                      M_V^2\left( \ln \frac{M_V^2}{\mu_0^2} + 2 \right)}
         \\[6mm]
         \multicolumn{2}{l}{\displaystyle K_1^{F_V}(\mu_0) = 
                   - \frac{3}{16}\frac{1}{16\pi^2}\frac{F_V^2}{F_0^2}
                   \left( \ln \frac{M_V^2}{\mu_0^2} 
                          + \frac{7}{6} \right)} 
         \\[8mm]
         K_2^{F_V}(\mu_0) = K_1^{F_V}(\mu_0) 
       & K_3^{F_V}(\mu_0) = - K_1^{F_V}(\mu_0)\\[2mm]  
         K_4^{F_V}(\mu_0) = 2 K_2^{F_V}(\mu_0) 
       & K_5^{F_V}(\mu_0) = - 3 K_1^{F_V}(\mu_0) \\[2mm]
         K_6^{F_V}(\mu_0) = - 3 K_2^{F_V}(\mu_0)\hspace{1cm}
       & K_{11}^{F_V}(\mu_0) = - K_1^{F_V}(\mu_0) \\[2mm]
         K_{12}^{F_V}(\mu_0) = 2 K_1^{F_V}(\mu_0) 
       & \\[2mm]
         K_{13}^{F_V} = \displaystyle
                        \frac{3}{4}\frac{1}{16\pi^2}\frac{F_V^2}{F_0^2} 
       & K_{14}^{F_V} = \displaystyle \frac{1}{2} K_{13}^{F_V}\;.\\[-2mm]
       & 
\end{array}
\end{eqnarray}
\begin{eqnarray}
\begin{array}{ll}
\underline{F_V G_V}: & \\[4mm]
         \multicolumn{2}{l}{\displaystyle K_1^{F_V G_V}(\mu_0) = 
         \frac{3}{4}\frac{1}{16\pi^2}\frac{F_V G_V}{F_0^2}
                         \left( \ln \frac{M_V^2}{\mu_0^2} 
                                + \frac{1}{6} \right)}
         \\[8mm]
         K_2^{F_V G_V}(\mu_0) = K_1^{F_V G_V}(\mu_0) 
       & K_3^{F_V G_V}(\mu_0) = - K_1^{F_V G_V}(\mu_0) \\[2mm] 
         K_4^{F_V G_V}(\mu_0) = 2 K_2^{F_V G_V}(\mu_0)
       & K_5^{F_V G_V}(\mu_0) = - 3 K_1^{F_V G_V}(\mu_0) \\[2mm]
         K_6^{F_V G_V}(\mu_0) = - 3 K_2^{F_V G_V}(\mu_0)\hspace{1cm}
       & K_{12}^{F_V G_V}(\mu_0) = K_1^{F_V G_V}(\mu_0) \\[2mm]
         K_{13}^{F_V G_V}(\mu_0) = - 2 K_1^{F_V G_V}(\mu_0)
       & K_{14}^{F_V G_V}(\mu_0) = K_1^{F_V G_V}(\mu_0)\;. \\[-2mm]
       & 
\end{array}
\end{eqnarray}
\begin{eqnarray}
\begin{array}{ll}
\underline{F_V^2 G_V^2}: & \\[4mm]
         \multicolumn{2}{l}{\displaystyle K_1^{F_V^2 G_V^2}(\mu_0) = 
         - \frac{3}{8}\frac{1}{16\pi^2}\frac{F_V^2 G_V^2}{F_0^4}
         \left( \ln \frac{M_V^2}{\mu_0^2} + \frac{7}{6} \right)}
         \\[8mm]
         K_2^{F_V^2 G_V^2}(\mu_0) = K_1^{F_V^2 G_V^2}(\mu_0) 
       & K_3^{F_V^2 G_V^2}(\mu_0) = - K_1^{F_V^2 G_V^2}(\mu_0) \\[2mm]
         K_4^{F_V^2 G_V^2}(\mu_0) = 2 K_2^{F_V^2 G_V^2}(\mu_0) 
       & K_5^{F_V^2 G_V^2}(\mu_0) = - 3 K_1^{F_V^2 G_V^2}(\mu_0) \\[2mm]
         K_6^{F_V^2 G_V^2}(\mu_0) = - 3 K_2^{F_V^2 G_V^2}(\mu_0)
         \hspace{1cm}
       & K_{13}^{F_V^2 G_V^2}(\mu_0) = - 4 K_1^{F_V^2 G_V^2}(\mu_0) 
         \\[2mm]
         K_{14}^{F_V^2 G_V^2}(\mu_0) = 2 K_1^{F_V^2 G_V^2}(\mu_0)\;. 
       &  \\[-2mm]
       & 
\end{array}
\end{eqnarray}
\begin{eqnarray}
\begin{array}{ll}
\underline{G_V^2}: & \\[4mm]
         \multicolumn{2}{l}{\displaystyle K_1^{G_V}(\mu_0) = 
         \frac{3}{8}\frac{1}{16\pi^2}\frac{G_V^2}{F_0^2}
         \left( \ln \frac{M_V^2}{\mu_0^2} + \frac{1}{6} \right)}
         \\[8mm] 
         K_2^{G_V}(\mu_0) = - K_1^{G_V}(\mu_0) 
       & K_3^{G_V}(\mu_0) = - K_1^{G_V}(\mu_0) \\[2mm] 
         K_4^{G_V}(\mu_0) =  2 K_2^{G_V}(\mu_0)
       & K_5^{G_V}(\mu_0) = - 3 K_1^{G_V}(\mu_0) \\[2mm] 
         K_6^{G_V}(\mu_0) = -3 K_2^{G_V}(\mu_0)\;.\hspace{1cm}
       &  \\[-2mm]
       & 
\end{array}
\end{eqnarray}
Using the relations $F_V G_V = F_0^2$ and $ F_V = 2 G_V$ (see
Eq.(\ref{sumrules})\,) the sum of the above expressions simplify to
\begin{eqnarray}\label{vector}
\begin{array}{ll}
         \displaystyle          
         Z^V(\mu_0) = - \frac{3}{16\pi^2}\frac{1}{F_0^2}M_V^2
                      \left( \ln \frac{M_V^2}{\mu_0^2} + 2 \right)
         \hspace{5mm}
       & \displaystyle 
         K_1^V(\mu_0) = \frac{3}{16}\frac{1}{16\pi^2}
                        \left( \ln \frac{M_V^2}{\mu_0^2} 
                               - \frac{23}{6} \right) \\[6mm]
         \displaystyle 
         K_2^V(\mu_0) = - \frac{3}{16}\frac{1}{16\pi^2}
                        \left( \ln \frac{M_V^2}{\mu_0^2} 
                               + \frac{25}{6} \right) 
       & K_3^V(\mu_0) = - K_1^V(\mu_0)\\[6mm] 
         K_4^V(\mu_0) = 2 K_2^V(\mu_0)  
       & K_5^V(\mu_0) = - 3 K_1^V(\mu_0) \\[6mm]
         K_6^V(\mu_0) = - 3 K_2^V(\mu_0)  
       & \displaystyle 
         K_{11}^V(\mu_0) = \frac{3}{8}\frac{1}{16\pi^2}
                        \left( \ln \frac{M_V^2}{\mu_0^2} 
                               + \frac{7}{6} \right) \\[6mm]
         \displaystyle 
         K_{12}^V = - \frac{3}{4}\frac{1}{16\pi^2} 
       & K_{13}^V = - 4 K_{12}^V\;. \\[-2mm]
       &
\end{array}
\end{eqnarray}
Note that the contributions to $K_{14}^V$ cancel in the sum.  
\vspace{5mm}

\underline{\em Axial vector Mesons}\hfill

\begin{eqnarray}\label{axialvector}
\begin{array}{ll}
         \multicolumn{2}{l}{\displaystyle 
         Z^A(\mu_0) = \frac{3}{2}\frac{1}{16\pi^2}\frac{F_A^2}{F_0^4}
                        M_A^2\left( \ln \frac{M_A^2}{\mu_0^2} + 2 \right)}
         \\[6mm]
         \multicolumn{2}{l}{\displaystyle
         K_1^A(\mu_0) = - \frac{3}{16}\frac{1}{16\pi^2}\frac{F_A^2}{F_0^2}
                        \left( \ln \frac{M_A^2}{\mu_0^2} 
                                 + \frac{7}{6} \right)} 
         \\[8mm]
         K_2^A(\mu_0) = - K_1^A(\mu_0) 
       & K_3^A(\mu_0) = - K_1^A(\mu_0) \\[2mm]
         K_4^A(\mu_0) =  2 K_2^A(\mu_0) \hspace{1cm} 
       & K_5^A(\mu_0) = - 3 K_1^A(\mu_0) \\[2mm] 
         K_6^A(\mu_0) = -3 K_2^A(\mu_0)
       & K_{11}^A(\mu_0) = K_1^A(\mu_0) \\[2mm]
         K_{12}^A(\mu_0) = 2 K_1^A(\mu_0)
       & \\[2mm] 
         \displaystyle
         K_{13}^A = - \frac{3}{4}\frac{1}{16\pi^2}\frac{F_A^2}{F_0^2} 
       & \displaystyle
         K_{14}^A = - \frac{1}{2} K_{13}^A \;. \\[-2mm]
       & 
\end{array}
\end{eqnarray}
With the help of the two Weinberg sum rules (see Eq.(\ref{sumrules})\,) 
the sum of the expressions from the vector resonances in (\ref{vector})
and from the axial vector resonances in (\ref{axialvector}) reads
\begin{eqnarray}\label{spin1}
\begin{array}{ll}
         \displaystyle 
         Z^{V+A} = \frac{3}{16\pi^2}\frac{1}{F_0^2}M_V^2 \ln 2 
       & \\[6mm]
         \displaystyle 
         K_1^{V+A} = - \frac{3}{16}\frac{1}{16\pi^2} (5 + \ln 2) 
       & \displaystyle 
         K_2^{V+A} = - \frac{3}{16}\frac{1}{16\pi^2} (3 - \ln 2) \\[6mm] 
         K_3^{V+A} = - K_1^{V+A} 
       & K_4^{V+A} = 2 K_2^{V+A}  \\[6mm] 
         K_5^{V+A} = - 3 K_1^{V+A} 
       & K_6^{V+A} = - 3 K_2^{V+A} \\[4mm] 
         \displaystyle 
         K_{11}^{V+A}(\mu_0) = \frac{3}{16}\frac{1}{16\pi^2}
                               \left( \ln \frac{M_V^2}{\mu_0^2} 
                                      + \frac{7}{6} - \ln 2 \right) 
       & \\[6mm] 
         \displaystyle 
         K_{12}^{V+A}(\mu_0) = - \frac{3}{8}\frac{1}{16\pi^2}
                               \left( \ln \frac{M_V^2}{\mu_0^2} 
                                      + \frac{19}{6} + \ln 2 \right) 
       & \\[6mm] 
         \displaystyle 
         K_{13}^{V+A} = \frac{9}{4}\frac{1}{16\pi^2}
       & \displaystyle
         K_{14}^{V+A} = \frac{1}{6} K_{13}^{V+A} \;. \\[2mm] 
       & 
\end{array}
\end{eqnarray}
Therefore only $K_{11}^{V+A}(\mu_0)$ and $K_{12}^{V+A}(\mu_0)$ are scale
dependent.
\vspace{5mm}

\underline{\em Scalar Mesons}\hfill

\begin{eqnarray}\label{scalar}
\begin{array}{ll}
\underline{{\rm Octet}}: & \\[4mm]
         \multicolumn{2}{l}{\displaystyle 
         K_1^S(\mu_0) = \frac{3}{4}\frac{1}{16\pi^2}\frac{c_d^2}{F_0^2}
                        \left( \ln \frac{M_S^2}{\mu_0^2} 
                        + \frac{1}{6} \right)} 
         \\[8mm]
         K_2^S(\mu_0) = - K_1^S(\mu_0) 
       & \displaystyle 
         K_3^S(\mu_0) = - \frac{1}{3} K_1^S(\mu_0) \\[2mm]
         \displaystyle 
         K_4^S(\mu_0) = - \frac{2}{3} K_1^S(\mu_0) \hspace{1cm}
       & K_5^S(\mu_0) = - K_1^S(\mu_0) \\[4mm]
         K_6^S(\mu_0) = K_1^S(\mu_0)\;. 
       & \\[-2mm]
       & 
\end{array}
\end{eqnarray}
\begin{eqnarray}
\begin{array}{ll}
\underline{{\rm Singlet}}:  & \\[4mm]
         \multicolumn{2}{l}{\displaystyle 
         K_3^{S_1}(\mu_0) =  - \frac{3}{2}\frac{1}{16\pi^2}
                             \frac{\tilde{c}_d^2}{F_0^2}
                             \left( \ln \frac{M_{S_1}^2}{\mu_0^2}
                             + \frac{1}{6} \right)}
         \\[8mm]
         K_4^{S_1}(\mu_0) = 2 K_3^{S_1}(\mu_0)\;. 
       & \\[-2mm]
       & 
\end{array}
\end{eqnarray} 
The numerical results at the scale point $\mu_0 = M_\rho$ (with the values
for the parameters given in Eqs.(\ref{values},\ref{scalarvalues})\,) are
presented in Table \ref{krnum}. There are no contributions to $K_7 \ldots
K_{10}$ from the resonances at all. For the nonvanishing couplings we
find that the contributions from the vector and axial vector resonances
dominate the $K^R_i$, like in the strong sector \cite{ecker89}.

\begin{table}[p]
$$
\begin{array}{|l|r|r|r|r|r|}
\hline &&&&& \\[-2mm]
\mbox{ Type: }   &    V   &    A   &    S   &    S_1  & \mbox{ Total } 
\\[2mm] \hline &&&&&\\[-2mm]
Z^R(M_\rho)      & - 0.88 & 1.79   & 0      & 0       & 0.91^{(*)}\\[2mm] 
\hline \multicolumn{6}{|c|}{} \\[-2mm]
\multicolumn{6}{|c|}{\mbox{ Units: } \times 10^{-3}} \\[2mm]
\hline &&&&&\\[-2mm]
K^R_1(M_\rho)    & - 4.6  & - 2.2  & 0.4    & 0       & - 6.4   \\[2mm] 
\hline &&&&&\\[-2mm]   
K^R_2(M_\rho)    & - 4.9  & 2.2    & - 0.4  & 0       & - 3.1   \\[2mm] 
\hline &&&&&\\[-2mm]
K^R_3(M_\rho)    & 4.6    & 2.2    & - 0.1  & - 0.2   & 6.4     \\[2mm] 
\hline &&&&&\\[-2mm]
K^R_4(M_\rho)    & - 9.9  & 4.4    & - 0.2  & - 0.5   & - 6.2   \\[2mm] 
\hline &&&&&\\[-2mm]
K^R_5(M_\rho)    & 13.7   & 6.6    & - 0.4  & 0       & 19.9    \\[2mm] 
\hline &&&&&\\[-2mm]
K^R_6(M_\rho)    & 14.8   & - 6.6  & 0.4    & 0       & 8.6     \\[2mm] 
\hline &&&&&\\[-2mm]
K^R_7 \ldots K^R_{10} & 0 & 0      & 0      & 0       & 0       \\[2mm]
\hline &&&&&\\[-2mm]
K^R_{11}(M_\rho) & 2.8    & - 2.2  & 0      & 0       & 0.6     \\[2mm] 
\hline &&&&&\\[-2mm]
K^R_{12}(M_\rho) & - 4.7  & - 4.4  & 0      & 0       & - 9.2   \\[2mm] 
\hline &&&&&\\[-2mm]
K^R_{13}         & 19.0^{(*)} & - 4.7^{(*)} & 0 & 0   & 14.2^{(*)}\\[2mm] 
\hline &&&&&\\[-2mm]
K^R_{14}         & 0      & 2.4^{(*)} & 0   & 0       & 2.4^{(*)} \\[2mm] 
\hline
\end{array}
$$
\caption[]{\label{krnum} Contributions from vector ($V$), axial vector
($A$), scalar octet ($S$) and scalar singlet ($S_1$) resonances to the
couplings $Z$ and $K_1 \dots K_{14}$ at the scale point $\mu_0 = M_V \simeq
M_\rho$. We have used the relations in Eq.(\ref{sumrules}) and the
numerical values for the parameters given in
Eqs.(\ref{values},\ref{scalarvalues}). Scale independent quantities
different from zero are signed with an asterisk ${}^{(*)}$.}
\end{table}

In Table \ref{silist} in Section \ref{res:section4} we list the algebraic
expressions for $K^R_1 \dots K^R_{13}$ ($K^R_{14}$ does not contribute to
physical amplitudes) using the relations in Eq.(\ref{sumrules}), also
indicated are the numerical results at the scale point $\mu_0 = M_\rho$.

\section{Large $N_C$}
\label{appendixnc}

In this appendix we classify the coupling constants $K_i$ according to
their large $N_C$ behaviour. We then compare their behaviour to the
numerical result obtained from the resonances.
  
In the large $N_C$ limit of QCD, operators that contain the product
of two traces in flavour space are suppressed by $1/N_C$ with respect to
those operators with one flavour trace only \cite{hooft74}. We apply this
counting scheme to operators of order $O(e^2 p^2)$ in the next-to-leading
Lagrangian ${\cal L}_4^Q$. First of all we have to make two assumptions:
\begin{enumerate}
\item We assume that the electromagnetic interactions do not change the
large $N_C$ counting of the strong force, i.e. 
\begin{equation}
M_{\pi}^2 \sim O(1), \hspace{1cm} F_{\pi} \sim O(N_C^{1/2}) \quad . 
\end{equation}
From this we deduce 
\begin{equation}
C \sim O(N_C), \hspace{1cm} K_i \sim O(1), \hspace{5mm}i = 1 \ldots 14
\quad ,
\end{equation}
where we made use of Eq.(\ref{cterm}) and of the expressions for the masses
\cite{urech95,neufeld95}, for the $P_{\ell 2}$ decay constants
\cite{neufeld95a} and for the vector and axial vector two point functions.
\item In the construction of ${\cal{L}}^C$ and ${\cal{L}}^Q_4$ local
spurions $Q_R(x), Q_L(x)$ have been used which at the end are identified
with the charge matrix $Q$ \cite{ecker89}. Therefore we consider $Q$ as a
(spurious) source term and assume that the operator $\langle Q^2 \rangle$
corresponds to a closed quark loop with a photon loop attached that 
interacts via gluons to other (flavourneutral) operators \cite{wyler96} .
\end{enumerate}
Therefore the following parts of the Lagrangian ${\cal L}_4^Q$ could be
suppressed, 
\begin{eqnarray}
\begin{array}{ll}
  K_{1} F^{2}_{0} \langle d^{\mu}U^{\dagger } d_{\mu}U \rangle 
                  \langle Q^{2} \rangle\;,
& K_{2} F^{2}_{0} \langle d^{\mu}U^{\dagger } d_{\mu}U \rangle 
                  \langle Q U Q U^{\dagger} \rangle \;,\\[2mm] 
\multicolumn{2}{l} 
  {K_{3} F^{2}_{0} \left( \langle d^{\mu}U^{\dagger } Q U \rangle 
                  \langle d_{\mu}U^{\dagger } Q U \rangle 
                  + \langle d^{\mu}U Q U^{\dagger} \rangle 
                  \langle d_{\mu}U Q U^{\dagger} \rangle \right)\;,}\\[2mm] 
  K_{4} F^{2}_{0} \langle d^{\mu}U^{\dagger } Q U \rangle 
                  \langle d_{\mu}U Q U^{\dagger} \rangle \;,
& \\[2mm]
K_{7} F^{2}_{0} \langle \chi^{\dagger} U + U^{\dagger} \chi \rangle 
                \langle Q^2 \rangle\;,
& K_{8} F^{2}_{0} \langle \chi^{\dagger} U + U^{\dagger} \chi \rangle
                  \langle Q U Q U^{\dagger} \rangle\;. 
\end{array}
\end{eqnarray}
However, in the derivation of ${\cal L}_4^Q$ we have used trace identities
that relate operators with different numbers of flavour traces
\cite{urech95}, which may invalidate the suppression of the single
operators listed above. Explicitly,
\begin{eqnarray}
\lefteqn{\langle U^\dagger d_\mu U Q U^\dagger d^\mu U Q 
+ U d_\mu U^\dagger Q U d^\mu U^\dagger Q \rangle = } \nonumber \\
&& 2 \langle \left\{ d^{\mu}U^{\dagger }, d_{\mu}U \right\} Q^2 \rangle 
   - \langle d^{\mu}U^{\dagger } d_{\mu}U \rangle \langle Q^{2} \rangle
\nonumber \\
&& + \left( \langle d^{\mu}U^{\dagger } Q U \rangle 
                     \langle d_{\mu}U^{\dagger } Q U \rangle 
                    + \langle d^{\mu}U Q U^{\dagger} \rangle 
                    \langle d_{\mu}U Q U^{\dagger} \rangle \right)\;,
\nonumber  \\ 
\lefteqn{\langle  d_\mu U^\dagger Q d^\mu U Q \rangle =} \nonumber \\
&& \langle d^{\mu}U^{\dagger } Q U \rangle 
   \langle d_{\mu}U Q U^{\dagger} \rangle 
   + \frac{1}{2}\langle d^{\mu}U^{\dagger } d_{\mu}U \rangle 
     \langle Q U Q U^{\dagger} \rangle  \nonumber \\ 
&& - \langle d^{\mu}U^{\dagger} d_{\mu}U Q U^{\dagger} Q U 
             + d^{\mu}U d_{\mu}U^{\dagger } Q U Q U^{\dagger} \rangle \;. 
\end{eqnarray}
Taking altogether, we find the following behaviour,
\begin{eqnarray}
K_i &\sim& O(1)\;, \hspace{1.5cm}i=1 \ldots 14;\; i \neq 7,8 \nonumber\\
K_1 + K_3,\;2 K_2 - K_4,\;K_7,\;K_8 &\sim& O(1/N_C) \;.
\end{eqnarray}
In comparison to the couplings $L_i$ in the strong sector the $K_i$ are in
general suppressed by a factor $1/N_C$ due to the additional factor $F_0^2$
split off in the electromagnetic counterterms (see Eq.(\ref{lag4})\,) and
due to the first assumption made above. The suppression of $K_8$ implies
that the contributions from the counterterms of order $O(e^2 m_s)$ to
$M_{\pi^\pm}^2$ are small,
\begin{eqnarray}
M_{\pi^\pm}^2 &=&  M_\pi^2 + 2 e^2 \frac{C}{F_0^2} 
                   + e^2 M_K^2 \left\{ 8 K^r_8(\mu) - \frac{C}{F_0^4} 
                   \left[ \frac{1}{8 \pi^2} \ln \frac{M_K^2}{\mu^2} 
                   + 32 L^r_4(\mu) \right] \right\} \nonumber \\
&& + O(e^2 M_\pi^2) \;.
\end{eqnarray} 
where $M_\pi, M_K$ are the leading order masses at $O(m_q)$ and the strong
coupling constant $L_4$ is suppressed itself in the large $N_C$ limit
\cite{gasser84}.

The contributions from the resonances to the couplings $K^R_i$ vary over a
large range, see Appendix \ref{appendixkr}. We look out for the numerical
contributions to the four quantities that are suppressed by $1/N_C$. $K_7$
and $K_8$ are untouched by the resonances, whereas $K^R_1 \ldots K^R_4$ get
rather large resonance contributions. The relevant combinations, however,
vanishes in nice agreement with the large $N_C$ limit,
\begin{equation}\label{nck2}
K^{R}_1(\mu_0) + K^{R}_3(\mu_0) =
2 K^{R}_2(\mu_0) - K^{R}_4(\mu_0) = 0 \quad .
\end{equation}


\end{document}